\documentclass[12pt,preprint]{aastex}

\newcommand{\ea}{et al.}

\newcommand{\kms}{\>{\rm km}\,{\rm s}^{-1}}

\newcommand{\myr}{\>{\rm Myr}}
\newcommand{\yr}{\>{\rm yr}}
\newcommand{\s}{\>{\rm s}}
\newcommand{\pc}{\>{\rm pc}}
\newcommand{\kpc}{\>{\rm kpc}}
\newcommand{\mpc}{\>{\rm Mpc}}

\newcommand{\ergsmic}{\>{\rm erg}\,{\rm cm}^{-2}\,{\rm s}^{-1}\,{\mu {\rm m}}^{-1}}

\newcommand{\mum}{\>{\mu {\rm m}}}
\newcommand{\kev}{\>{\rm keV}}
\newcommand{\ev}{\>{\rm eV}}

\newcommand{\msun}{\>{\rm M_{\odot}}}

\newcommand{\dg}{^{\circ}}

\newcommand{\as}{^{\prime\prime}}

\newcommand{\bdm}{\begin{displaymath}}
\newcommand{\edm}{\end{displaymath}}
\newcommand{\beq}{\begin{equation}}
\newcommand{\eeq}{\end{equation}}
\newcommand{\bit}{\begin{itemize}}
\newcommand{\eit}{\end{itemize}}
\newcommand{\ben}{\begin{enumerate}}
\newcommand{\een}{\end{enumerate}}
\newcommand{\bfi}{\begin{figure}[htb]}
\newcommand{\bpfi}{\begin{figure}[p]}

\newcommand{\htwo}{$\rm H_2$}
\newcommand{\fetwo}{$\rm [Fe\,II]$}

\newcommand{\brg}{$\rm Br\gamma$}
\newcommand{\ha}{$\rm H\alpha$}
\newcommand{\hel}{$\rm He\,I$}
\newcommand{\paa}{$\rm Pa\alpha$}

\newcommand{\htwoline}{$\rm H_2\,(1-0)\,S1$}

\slugcomment{submitted to AJ, July 27, 2007}

\shorttitle{Rings with SINFONI}
\shortauthors{B\"oker, T. et al. }

\begin{document}




\title{A SINFONI view of Galaxy Centers: 
Morphology and Kinematics of five Nuclear Star Formation Rings\altaffilmark{1}}

\author{Torsten B\"oker}
\affil{European Space Agency, Dept. RSSD, Keplerlaan 1, 2200 AG Noordwijk, Netherlands}
\email{tboeker@rssd.esa.int}

\author{Jes\'us Falc\'on-Barroso}
\affil{European Space Agency, Dept. RSSD, Keplerlaan 1, 2200 AG Noordwijk, Netherlands}
\email{jfalcon@rssd.esa.int}

\author{Eva Schinnerer}
\affil{Max-Planck-Institut f\"ur Astronomie, K\"onigstuhl 17, D-69117 Heidelberg, Germany}
\email{schinner@mpia.de}

\author{Johan H. Knapen}
\affil{Instituto de Astrof\'isica de Canarias, E-38200, La Laguna, Spain}
\email{jhk@iac.es}

\and
\author{Stuart Ryder}
\affil{Anglo-Australian Observatory, PO Box\,296, Epping, NSW\,1710, Australia}
\email{sdr@aao.gov.au}

\altaffiltext{1}{Based on observations collected at the European Southern 
Observatory, Chile, for proposals 076.B-0646(A) and 077B-0738(B).}

\begin{abstract}       
We present near-infrared (H- and K-band) integral-field observations of the
circumnuclear star formation rings in five nearby spiral galaxies. The data, obtained
at the {\it Very Large Telescope} with the SINFONI spectrograph, are used to
construct maps of various emission lines that reveal the individual star forming
regions ("hot spots") delineating the rings. We derive the morphological
parameters of the rings, and construct velocity fields of the stars  and the emission
line gas. We propose a qualitative, but robust, diagnostic for relative hot spot
ages based on the intensity ratios of the emission lines \brg , \hel , and \fetwo .
Application of this diagnostic to the data presented here provides tentative
support for a scenario in which star formation in the rings is triggered 
predominantly at two well-defined regions close to, and downstream from, the 
intersection of dust lanes along the bar with the inner Lindblad resonance.
\end{abstract}

\section{INTRODUCTION}
In many spiral galaxies of early- and intermediate Hubble type (Sa-Sc),  active
star formation is organized in a ring-like structure. These  star formation
rings offer a unique opportunity to study massive star formation in  external
galaxies: they produce up to 2/3 of the bolometric luminosity of their host 
galaxies \citep[e.g. NGC\,7469; ][]{gen95}, and often contain a large fraction
of the entire star formation activity of the galaxy.

The general picture of why molecular gas assembles in a ring is well understood
as a natural consequence of a non-axisymmetric gravitational  potential, nearly 
always due to the presence of a stellar bar or oval distortion 
\citep[e.g.][]{com85,kna95,hel96,but96}. Because of its dissipative nature, molecular
gas accumulates around the radii at which the stellar orbits experience
dynamical resonances with the rotating bar potential. Depending on the pattern
speed of the bar and the rotation curve of the galaxy, there can be one or
multiple such resonances. The high gas densities found in the rings, combined
with a variety of excitation mechanisms such as ultra-violet radiation from
young stars and mechanical shocks due to energetic outflows from massive stars
and/or an active galactic nucleus (AGN) help reveal the physical state
of the interstellar matter (ISM), be it molecular, atomic, or ionized gas.

Besides being fascinating laboratories in their own right, star formation rings
are important also for the secular evolution of disk galaxies \citep{kor04}. 
This is particularly true for the innermost of the dynamical resonances which
is called either the "inner Lindblad resonance" (ILR) or - in cases where a
compact massive  object leads to an additional dynamical resonance - the
``nuclear Lindblad resonance'' \citep*[NLR; ][]{fuk98}. Either of these
resonances can produce gas rings with radii of  a few hundred pc or less,
depending on the enclosed mass, and the rotation speed of the galaxy disk.  
On these spatial scales, a number of processes can cause the gas to lose  angular
momentum and to subsequently flow towards the nucleus. Examples for such
processes include torques due to the stellar potential \citep{gar05}, dynamical 
friction between giant molecular clouds that form within the ring due to self-gravity 
of the gas \citep*{fuk00}, the formation of spiral density waves \citep{eng00}, or mechanical energy released by star formation in the ring via stellar winds and/or supernova explosions.
 
Understanding the gas dynamics and star formation processes of 
nuclear\footnote{Throughout this paper, the term ``nuclear ring'' is used  to
imply that it is the innermost (star forming) gas ring that can be resolved
with the resolution of our data, typically a few hundred pc in diameter.} rings
in disk galaxies  is therefore crucial for developing models of gas
accumulation at the very nucleus of a galaxy and the evolution of any compact
massive object (CMO) which can exist in the form of a nuclear star cluster
\citep[NC,][]{boe02} and/or a  supermassive black hole (SMBH). While some
theoretical models exist on the gas behavior around a CMO
\citep[e.g.][]{fuk00}, observational data to constrain these models are rare,
mostly because of the limited  spatial resolution of mm-observations. Only
recently has it become possible to study  the molecular gas flows within a few
tens of pc from the nucleus in a small number of nearby  galaxies
\citep[e.g.][]{sch03,sch06,sch07}.

In order to increase the number of well-studied nuclear rings,  we have
begun a project to study the near-infrared (NIR) properties of five such
objects,  using the SINFONI integral-field spectrograph on the Very Large
Telescope (VLT). This paper discusses the morphologies, star  formation rates,
and kinematic properties of the rings. In a subsequent paper, we will make
use of the spectroscopic information contained in the SINFONI data  to 
investigate in more detail the star formation process, stellar populations, and
gas excitation  mechanism(s) in individual galaxies, both in the rings and the
galaxy nuclei.

This paper is structured as follows. In \S\ref{sec:data}, we describe the
sample selection, observational details, and data reduction techniques  common
to all galaxies. The continuum and emission line morphology of the rings as
well as the velocity fields of stars and gas for the individual galaxies are
presented in \S\ref{sec:diagnostics}. We discuss the results in the context
of competing models for the propagation of star formation in the rings in
\S\ref{sec:discuss}, and summarize our analysis in \S\ref{sec:summary}.

\section{OBSERVATIONS AND DATA REDUCTION}\label{sec:data}
\subsection{Galaxy Sample}\label{subsec:sample}
The general properties of the five galaxies discussed in this paper are
summarized in Table~\ref{tab:sample}. The objects were selected from  the
imaging survey of \cite{kna06} because they show evidence for small-scale
circumnuclear star formation rings within the SINFONI field of view, i.e. the 
central $8\as$. All five galaxies are classified as barred (SB) or weakly barred (SAB). 

In most cases, the ring is obvious in the \ha\ images of \cite{kna06} which
have a spatial resolution comparable to that of the SINFONI data discussed in
this paper.  The only exception is NGC\,1079 which requires HST resolution to
unambiguously identify the nuclear ring \citep{mao96}.

\subsection{SINFONI observations}\label{subsec:data}
SINFONI is an adaptive optics assisted, cryogenic 
NIR integral-field spectrograph, commissioned at the VLT.
It provides imaging spectroscopy of a contiguous, two-dimensional
field of $64\times64$ spatial pixels in the wavelength range from 
$1.1 - 2.45\mum$ at a resolving power of $2000 - 6000$. Details about the
instrument design can be found in \cite{eis03} and \cite{bon04} 

The bulk of the data described in this paper were obtained in service mode
during various nights between October 2005 and January 2006. 
One object (NGC\,5248) was observed
a semester later, in May/June 2006. The SINFONI instrument was used without
the adaptive optics module, and with its lowest magnification, resulting in
slice widths of $0.25\as$, a pixel scale of $\rm 0.125\as /pixel$, 
and a field-of-view of $8\as\times8\as$. This field
size is well-matched to the diameters of  the circumnuclear starburst rings
under study, hence no mosaicing was necessary. We used the SINFONI 
configuration for simultaneous H+K spectra which yields a spectral
resolution of R$\approx 2000$.

Because in all cases, the host galaxy is much more extended than the SINFONI
field of view, we obtained identical exposures of the  empty sky background by 
``nodding''
the telescope to point a few arcminutes  away from the galaxy nucleus. During
pipeline  processing, these ``sky frames'' are subtracted from the ``object
frames'', so that, at least in theory, the final data cubes are free of sky 
emission and dark current.

For each galaxy, we had requested an on-source integration time of 2.5\,hr,
divided into five observation blocks (OBs). Each OB consists of five $300\s$ 
long  on-source
exposures, paired with corresponding observations of the empty sky in an ABBA
nodding sequence. However, due to scheduling constraints, not all OBs were
actually executed. In some cases (e.g. NGC\,1079 and NGC\,1300), only one or
two OBs were executed. Although this  clearly reduced the signal-to-noise ratio
of the resulting data cubes, the emission line maps nevertheless yield
interesting insight into the structure of the rings. In Table~\ref{tab:obs} we
list the observation dates, OB identifiers, and the
resulting total integration times for all galaxies discussed in this paper.

\subsection{Data reduction}\label{subsec:reduc}
The reduced science products delivered by ESO for these datasets could not 
be used directly for our analysis. We identified two main reasons for this:

a) First, the default settings of the ESO pipeline uses the flag
{\sc objnod.scale\_sky = true} in the {\sc si\_rec\_objnod} recipe to
allow for scaling of the sky background before subtraction. This causes
unphysical (often negative) flux levels in the resulting data cubes, 
especially in regions where the sky flux level is significant compared
to the science object. It is thus important that this option is switched 
off when running the SINFONI pipeline on this type of raw data.

b) Second, in the raw, two-dimensional detector frames, the sky emission lines 
show a small, but noticable shift in their position. This misalignment
between sky and object frames causes imperfect removal of the sky lines 
in the resulting spectra, and is probably due to flexure in the optical
train. 

While the first issue could be easily solved, the proper removal of the sky
residuals required a more elaborate data reduction process. We used the
ESO pipeline for SINFONI (version 1.3) to perform the basic data reduction 
(bias subtraction, flatfielding, wavelength calibration and extraction)
of the individual frame pairs in each OB. For each pair, we improved on the 
sky subtraction by means of an IDL\footnote{http://www.ittvis.com/}
software package described in \cite{dav07} and kindly provided to us by R. Davies.
The algorithm allows to separately scale different transition groups of 
telluric OH lines before subtraction which minimizes the residuals in the 
final spectra. We aligned all OBs to be centered on the galaxy nucleus (i.e. the peak of the continuum emission), thus accounting for telescope dithers between OBs.
Moreover, we shifted all individual channel maps within an OB to correct
the atmospheric refraction which causes small shifts with wavelength of the 
continuum peak within an exposure. Finally, we median-combined the aligned 
and sky-subtracted OB cubes of a given galaxy.

The resulting ``final'' data  cube of each OB contains about 2100 channel maps
of the ($8\as\times8\as$) field of view with a spatial sampling of 
$0.125\as$/pixel and a spectral resolution of R$\approx 2000$. The OB
cubes are wavelength-calibrated and sky-subtracted, but at this stage neither
flux-calibrated nor corrected for the absorption features introduced by the
Earth's atmosphere. In order to correct for telluric absorption, we divided
each data cube by an atmospheric transmission model generated with the ATRAN
package \citep{lor92}. We created model atmospheres  for the altitude of the
Paranal observatory and for a number of zenith  angles spanning the range of
our observations ($0\deg - 30\deg$).

For flux calibration of the cubes, we used observations of  early-type (around
B5V) standard stars which are part of the routine  nightly SINFONI calibration.
For each night, we extracted the  observed stellar spectrum (in counts/s) in a
$5\as$ aperture and - after  correcting for telluric absorption as described
below - divided it into  the ``true'' intrinsic spectrum of the respective
spectral type. This,  in principle, yields the correction ``spectrum'' that
must be multiplied into the data cubes obtained during the same night in order
to obtain  the correct SED of the galaxy.  The ``true'' stellar spectrum was
obtained from the library of \cite{pic98} after scaling to the 
near-infrared magnitudes listed in the Hipparcos catalog\footnote{using the zero
points of the UKIRT photometric system of $1.12\cdot 10^{-6} (4.07\cdot
10^{-7})\ergsmic$ for  H-band (K-band).}. For some stars, the exact spectral 
type was not available in the Pickles library, in which case we used the 
nearest available type (the difference was never more than one subclass). 

Performing this flux calibration on a night-by-night basis produced a set of
galaxy SEDs that showed a relatively large scatter of about 15\% rms. 
This scatter is larger than that achieved by using a single calibration
spectrum for all OBs, regardless of when they were taken. This demonstrates
that the accuracy of the flux calibration is limited by  the systematic
uncertainty of the ``true'' spectra of the standard stars (and possibly 
by the accuracy of the telluric correction). We therefore decided to apply 
the same {\it average}
calibration spectrum to all individual OBs. This average calibration spectrum
was obtained by fitting a fourth-order polynomial to the average inferred
correction spectra of all standard stars. 

The systematic uncertainty of this approach should not be larger than the 
night-to-night variations of the flux calibration. We therefore estimate that
our flux calibration is accurate to 15\%. As a consistency check,
we compared the continuum flux levels of the calibrated spectra
in H- and K-band to those derived by the 2MASS survey. 
The agreement was better than 10\%, thus confirming our error estimate. 

As a last step in the data reduction, and after the telluric correction and 
flux calibration was applied to the individual OBs, they were merged to 
produce a fully reduced datacube for each galaxy.

\section{Near-Infrared Diagnostics: Stars, Gas, and Kinematics}\label{sec:diagnostics}
In this section, we present - for each of the five galaxies - a number of  maps
extracted from the SINFONI data cubes following the data reduction  process
just described. These maps contain a large amount of morphological and
kinematic information on the stars and various components of the ISM 
(molecular and ionized gas, dust, etc.). The various emission lines, in
particular,  trace different components of the inter-stellar medium (ISM) and
different excitation  mechanisms so that even a qualitative comparison provides
insight  into the current physical conditions inside the starburst rings.

\begin{figure}
\begin{center}
\includegraphics[angle=0,scale=.7]{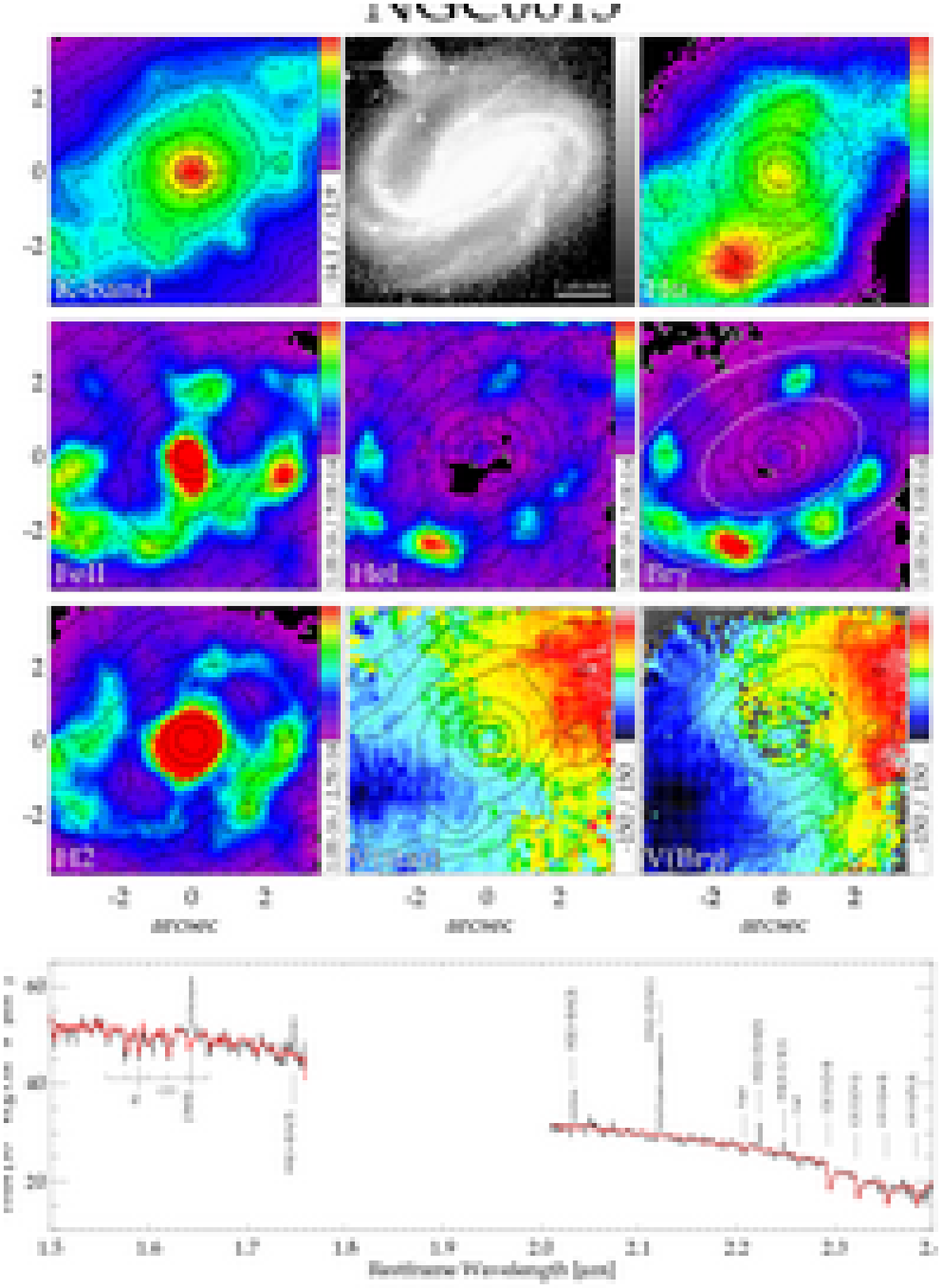}
\caption{Near-infrared morphology of NGC\,613. Please see 
\S~\ref{sec:diagnostics} for a description of the individual 
maps. Black contours delineate the K-band continuum from
the top-left panel. The white ellipses in the \brg\ map
mark the ring aperture used for the line flux measurements listed
in Table~\ref{tab:flux}.
In all panels, north is up and east is to the left.}
\label{fig:n613}
\end{center}
\end{figure}

\begin{figure}
\begin{center}
\includegraphics[angle=0,scale=.7]{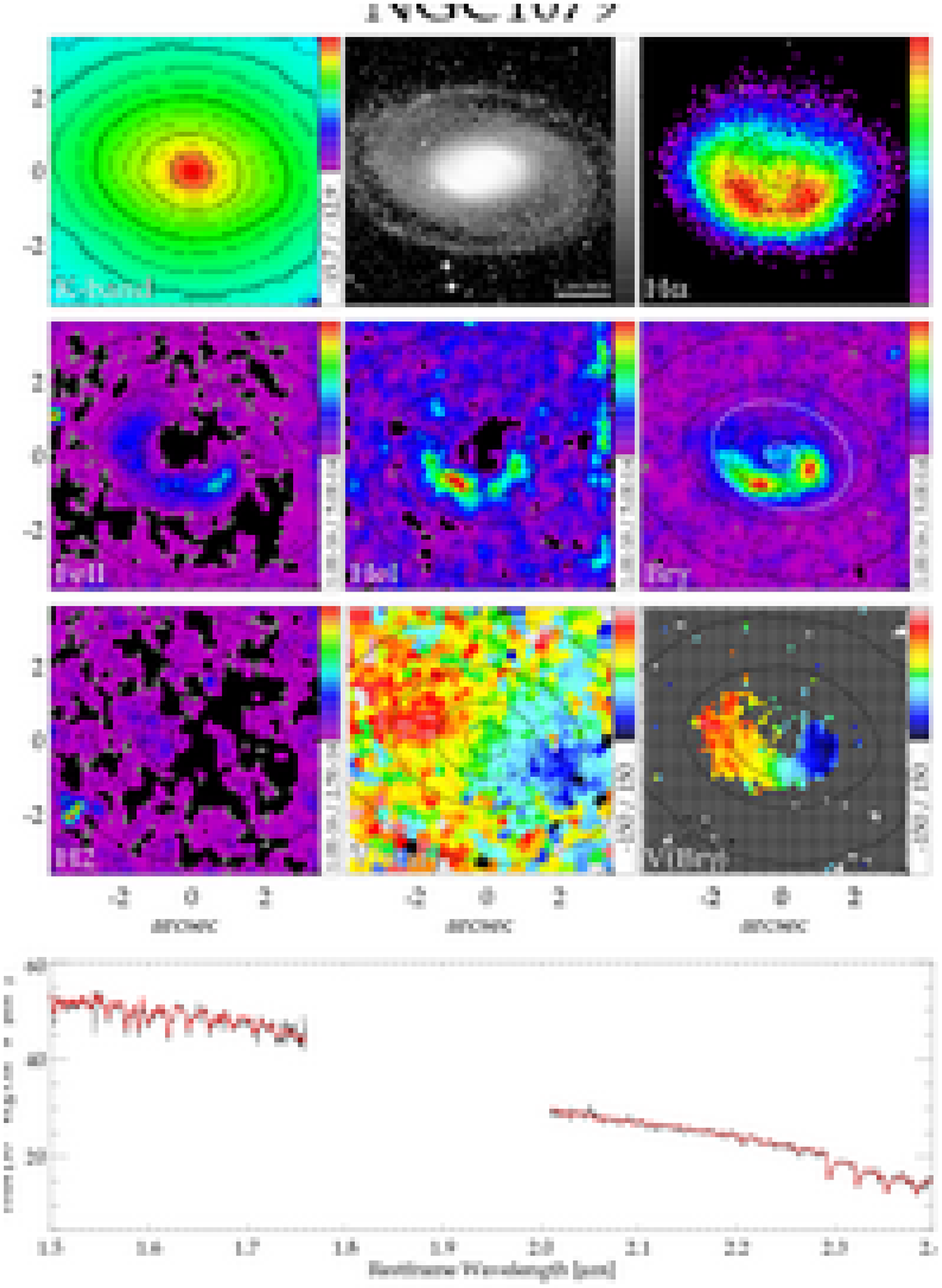}
\caption{As Figure~\ref{fig:n613}, but for NGC\,1079.
See Figure~\ref{fig:n613} for identification of spectral features.}
\label{fig:n1079}
\end{center}
\end{figure}

\begin{figure}
\begin{center}
\includegraphics[angle=0,scale=.7]{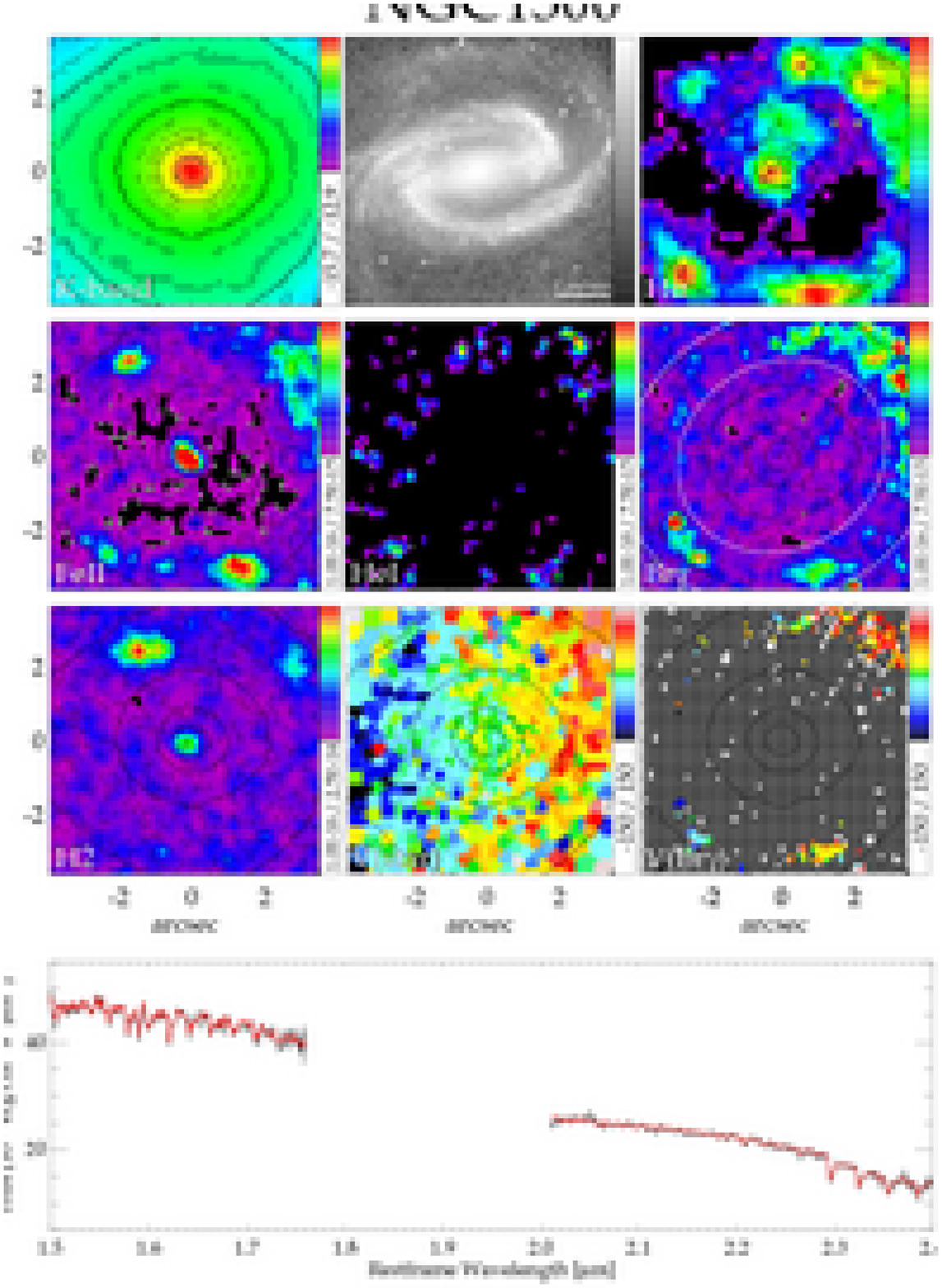}
\caption{As Figure~\ref{fig:n613}, but for NGC\,1300.
See Figure~\ref{fig:n613} for identification of spectral features.}
\label{fig:n1300}
\end{center}
\end{figure}

\begin{figure}
\begin{center}
\includegraphics[angle=0,scale=.7]{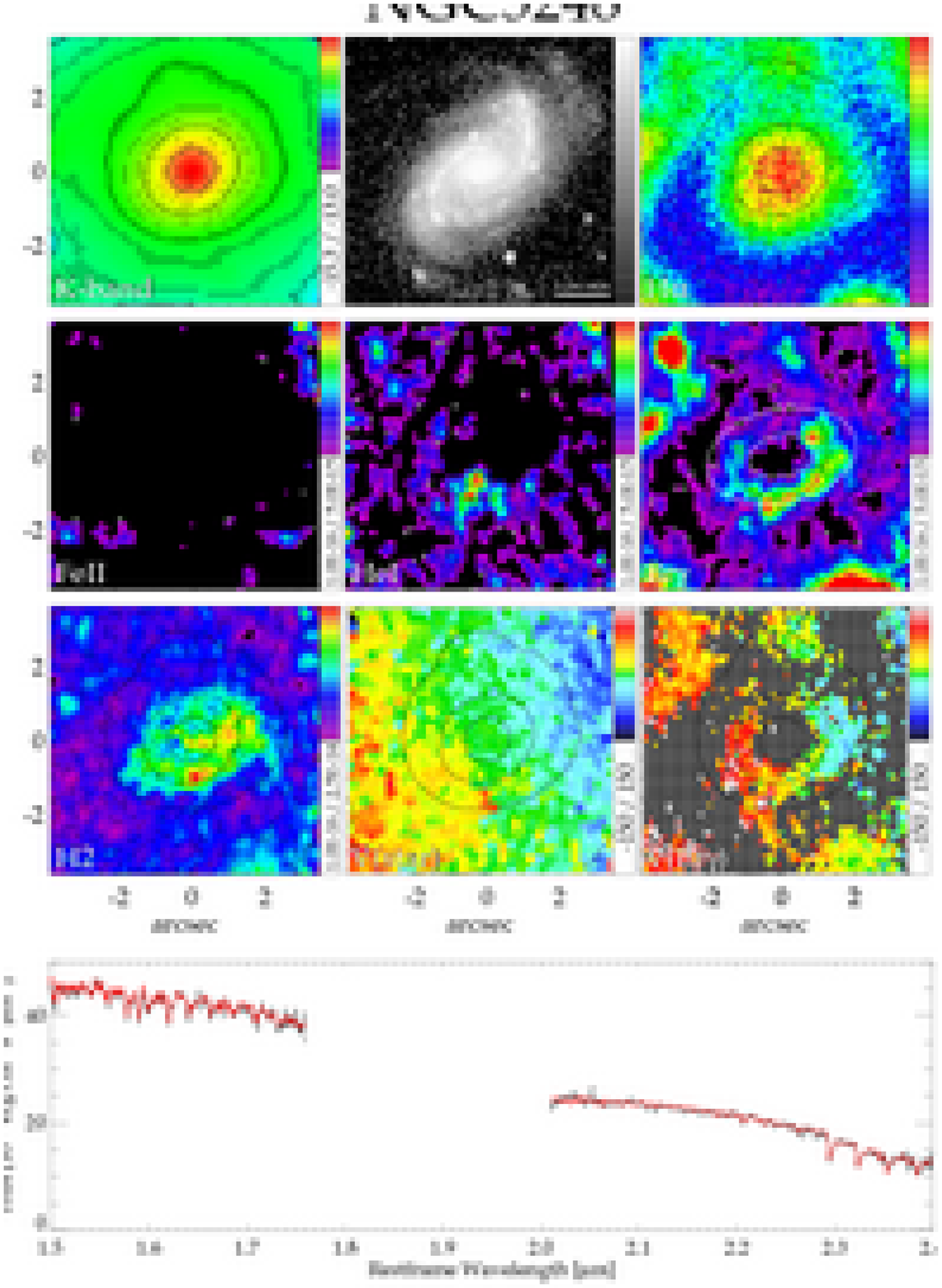}
\caption{As Figure~\ref{fig:n613}, but for NGC\,5248.
See Figure~\ref{fig:n613} for identification of spectral features.}
\label{fig:n5248}
\end{center}
\end{figure}

\begin{figure}
\begin{center}
\includegraphics[angle=0,scale=.7]{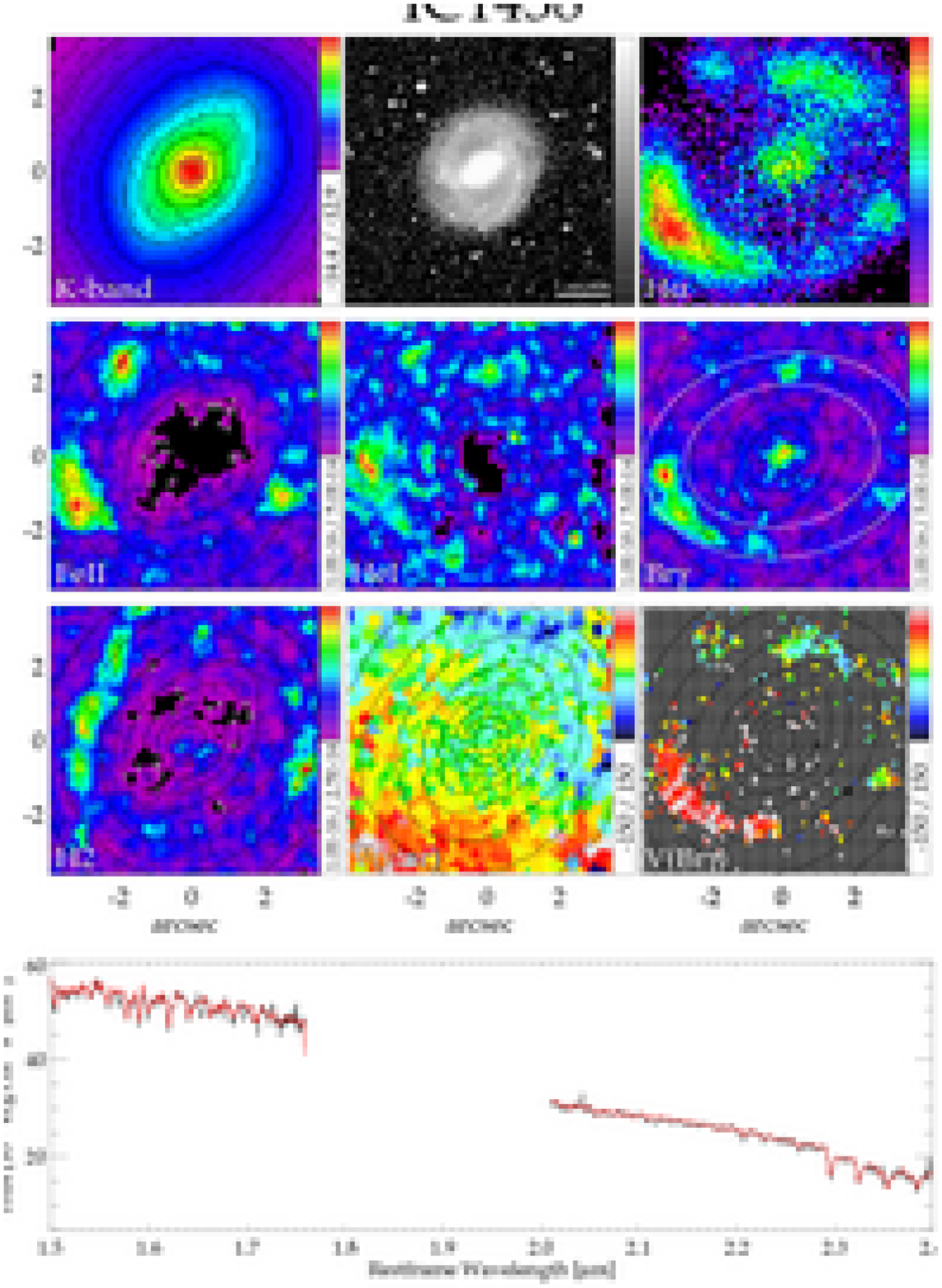}
\caption{As Figure~\ref{fig:n613}, but for IC\,1438.
See Figure~\ref{fig:n613} for identification of spectral features.}
\label{fig:i1438}
\end{center}
\end{figure}

Before we discuss each galaxy in more detail, we briefly summarize how each of 
the maps presented in Figures~\ref{fig:n613} to \ref{fig:i1438} (from left to
right and top to bottom) was constructed, what they signify, and how they can be
interpreted:

\ben

\item{A map of the K-band continuum emission. In the NIR
wavelength range, the continuum emission is predominantly produced 
in red giants and supergiants, and the mass-to-light ratio 
of evolved stellar populations is nearly independent of stellar type.
Therefore, the K-band morphology is an accurate tracer of the gravitational 
potential. Our K-band maps maps are constructed by simply adding up all 
channel maps in the
wavelength range $2.0 - 2.4\mum$, excluding those expected to contain
strong line emission, e.g. from \brg\ or \htwo . The result should be a
``clean'' map of the stellar mass distribution.}

\item{A $5\times5$ arcmin large-scale $R$-band image of the galaxy from the 
Digital Sky Survey.}

\item{A map of \ha\ emission in the central  $8\as\times8\as$ taken
from \cite{kna06}. We include these maps here to allow a visual comparison
to the SINFONI line maps, in particular \brg . However, the comparison should
be made keeping in mind some complications. First, the \ha\ images in \cite{kna06}
were constructed by subtracting a scaled I-band image, used as a substitute for
narrow-band continuum images. As discussed in \cite{boe99}, this method can cause 
imperfect continuum subtraction in regions with large color deviations, in particular 
the galaxy nucleus.
As can be seen by comparison to the SINFONI \brg\ maps (which have a more robust
continuum subtraction, see below), this is indeed the case for most of our sample.
Secondly, \ha\ is potentially more affected by dust extinction than \brg . Finally, the 
spatial resolution of the \ha\ maps is somewhat lower than that of the SINFONI data.
Nevertheless, these \ha\ maps provide a valuable consistency check of our NIR line maps.
Here, and in each of the following maps, we overplotted the K-band continuum contours
from panel 1 for comparison.}

\item{A \fetwo\ line map. The strongest emission line of \fetwo\ in the H- and
K-bands  is the $\rm a^4F_{9/2}-a^4D_{7/2}$ transition at $1.644\mum$. The most
likely excitation mechanism is collisions with free electrons in hot gas with
temperatures in the range $\rm 10^3-10^4\>K$. However, the ionization 
potential of \fetwo\ is only 16.2\,eV. This means that in regions of fully
ionized  hydrogen, most of the Fe atoms are in higher ionization states.
Therefore,  the \fetwo\ emission is not expected to be strong in normal HII
regions \citep*{mou00}. Instead, it will be strong in regions that i) have a
high abundance of Fe in the gas phase, and ii) are only partially ionized. The
two most likely mechanisms to create  extended partially ionized zones are
power-law photoionization by X-rays (e.g. from  an AGN) and fast shocks
produced in supernova remnants \citep{lab06}.}

\item{A \hel\ line map, more specifically a map of the $(n=2)\,^1P-^1S$ 
transition at $2.058\mum$ which is the strongest \hel\ line in
the H- and K-bands. The ionization energy of atomic helium is 
$24.6\ev$ and thus markedly higher than that of hydrogen ($13.6\ev$). Line
emission from \hel\ is  therefore expected to arise predominantly in the
vicinity of the most massive (and hence youngest) stars.  
Interpretation of the \hel\ emission  requires detailed
photo-ionization models, and quantitative analysis of the temperature of the
hottest stars in an H\,II region is subject to large uncertainties
\citep{doh95,lum01}. At least qualitatively, however, comparing the \hel\ line
strength  to that of \brg\ provides a handle on the {\it relative} ages of the
youngest  stellar clusters because the hottest stars will vanish fastest. We
will use this  simple diagnostic in the discussion of age gradients along the
rings in \S\ref{subsec:ages}.}

\item{A map of the \brg\ line at $2.16\mum$. 
The \brg\ recombination line of the hydrogen atom is
predominantly  produced by photo-ionization in the Str\"omgren-spheres around
O- or B-type stars. Although fast shock-fronts due to supernovae and/or strong
stellar winds can also ionize hydrogen atoms, these mechanisms are related to
the same population of OB stars, and thus they occur in the same
locations\footnote{It is, in principle, possible to distinguish between
ultra-violet photons and fast shocks as the cause of hydrogen ionization: the
former produces low HII densities with a large filling factor (i.e. high
column densities), while the latter produces high densities with a low filling
factor.}.  A third potential ionization mechanism is strong non-thermal
emission from an active galactic nucleus (AGN). Outside the galaxy nucleus,
however, we consider \brg\ as a primary tracer of young stellar populations. 

The \brg\ map, as well as the other emission line maps presented in
Figures~\ref{fig:n613} to \ref{fig:i1438} was constructed by summing up all
channel maps containing line emission, and subtracting from each a  continuum
map created by averaging five channel maps on either side of the line. This is
equivalent to the standard method of taking  narrow-band images through filters
that are centered on the line and the red and blue continuum, respectively.}

\item{A \htwo\ line map. The (1-0)\,S1 transition of the \htwo\ molecule at
$2.122\mum$ is a tracer of warm molecular gas with temperatures around $\rm
\approx 2000\>K$. There are three plausible mechanisms that can heat molecular 
gas to these temperatures:
i) UV-fluorescence in photo-dissociation regions (PDRs), typically
the surface layers of dense molecular clouds in the vicinity of UV-bright star
forming regions \citep[e.g.][]{bla87}, ii) collisional excitation of the \htwo\
molecules in fast shock fronts \citep[e.g.][]{hol89}, and iii) X-ray illumination
\citep*{mal96}. Using the relative strengths of the numerous other rotational 
and vibrational transitions of the \htwo\ molecule present in the NIR spectral 
range, one can, in principle, distinguish between these excitation modes.
In the case of NGC\,613, we will present this analysis in a forthcoming paper
\citep[][hereafter paper II]{fal08}. However, for the
purpose of this paper, the \htwoline\ line maps merely serve to indicate the
distribution of warm molecular gas. }

\item{A map of the stellar velocity field, derived using the penalized 
pixel-fitting (pPXF) method of \citet{capem04}. We made use of a subset of the
library of theoretical spectra from \citet{lancon07} adapted to our spectral
resolution. This subset comprises
stellar  models with solar abundances for a range of effective temparature
(T=2900-5900 K)  and gravity (log(g)=0-2). In order to ensure reliable measurements
of stellar  kinematics, we spatially binned our datacubes using the
Voronoi 2D binning algorithm of \citet{michele03}, creating compact bins with a
minimum signal-to-noise ratio ($S/N$) of $\sim30$ per spectral resolution
element.

We fitted a non-negative linear combination of these models, convolved with a Gaussian
line-of-sight velocity distribution (LOSVD), to the spectrum of each individual 
pixel over a wavelength range centered on the $2.29\mum$ CO bands, masking 
any potential emission lines. The fit also allows for a low-order 
Legendre polynomial in order to account for small differences in
the continuum slope between the pixel spectra and the stellar templates.
The best-fitting template mix was then determined by 
$\chi^2$ minimization in pixel space. }
	
\item{A map of the \brg\ velocity field, determined using the 
GANDALF code \citep{sarzi06}. This routine fits simultaneously both the 
stellar continuum and  emission lines, treating the emission lines as 
additional Gaussian templates. In order to subtract the stellar continuum, 
we used the best template combination from the stellar kinematic fit described 
above. In general, the high signal-to-noise ratio of \brg\ allows to reliably 
measure the gas kinematics in smaller spatial bins, which yields higher spatial
resolution than for the stellar kinematics, especially in the outer regions. 
However, the \brg\ kinematics can only be derived where the emission line
is strong. In the figures, we only display velocities for regions with a S/N 
above 2, while regions below this S/N are left blank (gray areas).}

\item{Finally, we present the H- and K-band spectrum of the galaxy
nucleus, i.e. a circular aperture with $1\as$ diameter centered on
the K-band peak.
Overplotted in red is the best stellar template fit from item 8. 
In the case of NGC\,613 (Fig.~\ref{fig:n613}), the emission lines 
as well as some prominent stellar absorption features are labeled
for convenience.}

\een

The quantitative analysis of the line fluxes over the entire SINFONI field as
well as within an aperture encompassing the ring (as delineated in each of the
\brg\ maps) is summarized in Table~\ref{tab:flux}. In what follows, we comment
in more detail on the individual galaxies. In particular, we describe those
properties of the circumnuclear starburst rings that can be deduced from the
images described above. 

\subsection{NGC\,613} 
NGC\,613 is a large, relatively nearby ($d\,=\,17.5\mpc$), strongly barred  Sbc
spiral. It is classified as a Seyfert galaxy in the NED database,  although
without a type designation, and in fact, it is somewhat unclear  on what basis
this classification was made.  The optical emission line spectrum of the
NGC\,613 nucleus is  classified as ``composite'' by \cite{ver86}. Furthermore,
NGC\,613 emits neither detectable X-ray emission  in the $2-10\kev$ band, nor
any high excitation MIR lines such as [OIV] or [SiII] that are usually found in
AGN \citep*{sat04}. In \S\ref{subsec:nuclei}, we discuss that the nucleus is
also quiescent  in the NIR, in the sense that it does not ionize much of the
surrounding gas. On the other hand, radio continuum observations show evidence
for  an energetic outflow which has been interpreted as a jet emanating from
an  AGN \citep{hum87,hum92}. The same radio observations also clearly reveal
the star formation ring.

The excellent signal-to-noise ratio of our SINFONI data for NGC\,613 allows a 
detailed spectroscopic study of the star formation processes and stellar 
populations in the nucleus and nuclear ring of NGC\,613 which we present 
in paper II. Here, we highlight a few interesting characteristics that are 
evident from the NIR maps alone.

\subsubsection{Continuum emission} 
There is some dispute in the
literature about the shape of the gravitational potential in NGC\,613. On the
one hand, \cite*{jun97} and \cite{esk02} have suggested that NGC\,613 has a
nuclear bar at the same position angle (PA) as the  large-scale bar \citep[$110\dg$;
][]{hum87}. However, based on HST imaging,  \cite{erw04} raised the possibility
that the ``nuclear bar'' was in fact  a spurious feature caused by the presence
of an elliptical ring of  star clusters. As can be seen from the following
discussion, such a ring is indeed present, and the morphology of our K-band map
in Figure~\ref{fig:n613} is clearly affected by the continuum emission from the ring.
For this reason, and because of the limited SINFONI field of view 
we cannot decide whether the galaxy morphology underlying 
the ring shows a bar-shaped stellar mass distribution. 

\subsubsection{Emission line morphology - evidence for an AGN outflow}\label{subsec:outflow} 
The \brg\ emission shows a very clearly defined ring structure, composed  of seven
almost regularly spaced bright clumps that are the sites of current massive
star formation. These ``hot spots'' are brightest along the southern half of
the ring, while the northern half shows a well-defined ``gap'' at PA $30\dg$
which is also  evident in the \fetwo\ emission map. This direction agrees well
with that of the radio  jet found by \cite{hum87} and studied further by
\cite{hum92}. It thus appears that a mechanical outflow from the AGN has
disturbed the ring morphology. The outflow thus must intersect the plane of
the disk. 

\begin{figure}
\begin{center}
\includegraphics[angle=0,scale=.8]{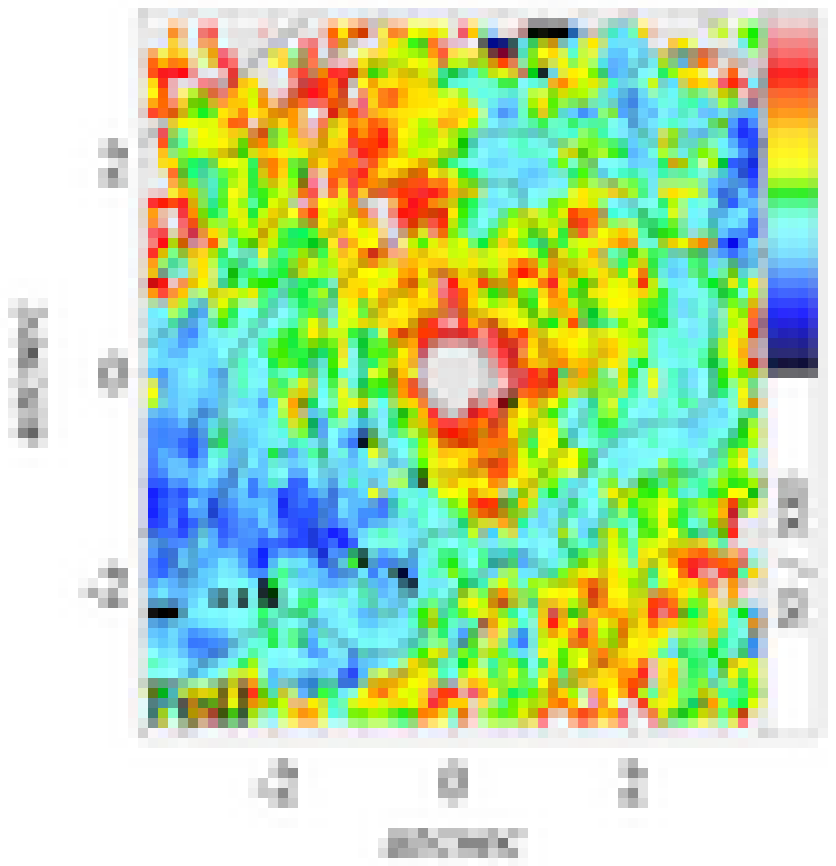}
\caption{Dispersion map of the \fetwo\ emission line in NGC\,613.
Note the line broadening along the radio jet at PA $30\dg$ \citep{hum92}}
\label{fig:sigma}
\end{center}
\end{figure}

Additional evidence for such an outflow comes from the \ha\ image of
\cite{kna06} which shows two faint``streamers'' that are closely 
aligned with the radio jet. 
Moreover, a map of the velocity dispersion of the ionized gas (as
measured by the line width of \fetwo ) shown in Figure~\ref{fig:sigma}
clearly shows abnormally high dispersions along both sides of the outflow. 
The influence of the radio jet on the nuclear ring in NGC\,613
will be addressed in more detail in paper II.

The \htwo\ emission also shows the nuclear ring, but is strongest in
the  nucleus. The \htwo\ ring is not composed of distinct hot spots like the
\brg\ emission,  but is more smoothly distributed. There are two emission peaks,
found on opposite  sides of the ring, at approximate PA of $90\dg$ and
$270\dg$. If the \htwo\ emission in the ring of NGC\,613 were purely caused
by UV radiation, one might expect a spatial correspondence between \htwo\
and the ionized gas traced by \brg\ and/or \hel . However, many of the \brg\ hot spots
are not bright in \htwo . Therefore, the \htwo\ emission probably contains a
non-negligible contribution from shock-heated molecular gas.

\subsection{NGC\,1079} 
NGC\,1079 is an isolated, grand-design spiral in the Eridanus supergroup
\citep{bro06} marked by a bright bulge and very faint outer disk. It represents
a good example for a galaxy in which nearly the entire star formation activity
takes place in the ring, as can be seen by the \ha\ image of \cite{kna06} which
shows an extremely quiescent galaxy disk

The nuclear ring of NGC\,1079 has first been identified by \cite{mao96} from
HST ultraviolet imaging and ground-based spectroscopy which confirmed that the
gas rotation curve is consistent with a rotating ring.

\subsubsection{Continuum emission}
Within the rather small SINFONI field of view, the K-band map of NGC\,1079 
appears smooth and featureless, in agreement with the H-band map of \cite{jun97}. 
Within the central $8\as$, the K-band isophotes actually appear to have 
a somewhat smaller P.A. of $\approx 75\dg$, but this is also evident in 
the \cite{jun97} map. These authors find evidence for a secondary bar 
with P.A. $96\dg$ and major axis length of $17\as$. It is possible that 
this bar governs the gas dynamics in the inner few hundred pc. 
Unfortunately, it is not possible to determine the dust structure
in NGC\,1079 because no high-resolution optical/NIR imagery is available 
in the HST archive, except for the UV data of \cite{mao96} which are 
however not suitable to derive extinction variations.

\subsubsection{Emission lines}
Despite its small angular size (radius $\approx 1.5\as$), the star formation
ring in NGC\,1079 is easily resolved in our SINFONI maps. The line maps of
\brg , \htwo , and \fetwo\ all show strong emission from the southern half of
the ring, while the northern half is either quiescent or deeply obscured. The
\brg\ map shows two hot spots of roughly equal flux at P.A. $\approx 150\dg$ 
and $240\dg$. Both are also seen in \hel , but in this line the eastern
hot spot is substantially stronger. 
We detect little or no emission from \htwo\ at the sensitivity limit of our 
data (note that the data set for NGC\,1079 only consists of two OBs, i.e.
only one hour on-source integration time).

\subsection{NGC\,1300} 
The SBbc spiral NGC\,1300 is the prototype of a strongly barred disk galaxy, 
and has been extensively used as a testbed for dynamical modelling 
\citep{lin96,lin97,agu00}. The dust structure in the central kpc as revealed by
the V-H color map  of \cite{mar03} does not resemble a ring, but rather is a
good example of a grand design spiral. 

NGC\,1300 shows no signs of nuclear activity, and \cite{sca04} find no
evidence for a distinct stellar nucleus. Nevertheless, from an HST/STIS 
study of the kinematics of the nuclear emission line gas, \cite{atk05} 
conclude that NGC\,1300 hosts a SMBH of $6.6\cdot10^6\msun$. 

\subsubsection{Continuum emission}
As in the case of NGC\,1079, the continuum morphology of NGC\,1300 is very
smooth, which has also been noted by \cite{per00}. 
There is no signature of the ring evident,
indicating that in terms of stellar mass, the star formation ring is a small
perturbation to the overall potential. 
 
\subsubsection{Emission lines}
Of all galaxies in our sample, NGC\,1300 is the one with the least integration time,
and the SNR of the line maps, and in particular the velocity fields is therefore
rather low. Nevertheless, the ring structure is clearly apparent, especially in
\brg\ and \fetwo . Both lines show a number of individual hot spots, with the two
brightest \brg\ clumps located at opposite ends of the ring along its major axis
(P.A. $\approx 125\dg$ and $305\dg$). In contrast, the two strongest \fetwo\ clumps 
are located along the ring minor axis (P.A. $\approx 30\dg$ and $210\dg$). The northern
\fetwo\ clump is also bright in \htwo .

\subsection{NGC\,5248} 
The grand-design Sbc spiral NGC\,5248 is a ``posterchild'' example for the
impact of a stellar bar on the gas dynamics and circumnuclear star formation. 
\cite{jog02a} have convincingly argued that the spiral structure is being
driven by an extended, moderately strong stellar bar that has a  deprojected
ellipticity of 0.44 and a semimajor axis $a_{bar}$ of $95\as$ ($7.1\kpc$). The
inner kiloparsec of NGC\,5248 has been well studied with ground-based images
in emission lines \citep{ken89,lai01,kna06} as well as optical \citep*{kna06}
and NIR continuum \citep{elm97,per00}.  These images have revealed a $\approx
750\pc$ diameter star formation ring which, when seen at HST resolution, is
clearly resolved into numerous bright HII regions and young star
clusters \citep{mao96,mao01}. 

Near-infrared color maps have also revealed a nuclear dust spiral \citep{lai99}
that spans the radial range between $1\as (75\pc)$ and $4\as (300\pc)$. This
structure likely forms the inward continuation of the grand-design  spiral arm
system that is apparent in various tracers out to a radius of about $10\kpc$
\citep{jog02a}. At its inside, the nuclear dust spiral appears to lead into a 
second \ha -bright ring that has a radius of only $1.25\as\ (100\pc)$
\citep{lai01,mao01} and thus forms one of the smallest such structures known.

As with NGC\,613, NGC\,5248 is believed to harbor an AGN, and is
classified as Sy\,2. However, at the sensitivity limits of the SINFONI data,
we find no evidence for emission lines in the NIR spectrum of the NGC\,5248 nucleus.

\subsubsection{Continuum emission}
The K-band morphology in the center of NGC\,5248 has been discussed
in detail by \cite{jog02a}. Our SINFONI map agrees well with the
one presented in their Figure~8. In particular, both the ``very weak
oval feature of radius $3\as$'' and the excess flux north of the nucleus 
are present in the two maps. As discussed by \cite{jog02a}, the weak oval could 
be due to a disk-like component, a late-type bulge, or an unresolved nuclear bar. 
Our data cannot further constrain its origin, but we note that the stellar
velocity field shows rather regular rotation over the SINFONI field, without
any signs for kinematically distinct components. 

\subsubsection{Emission lines}
The two nested star formation rings of NGC\,5248 can be seen in the 
SINFONI \brg\ map. However, only the inner edge of the larger ring fall within
the SINFONI field. The map clearly resolves the inner ring of NGC\,5248.
The morphology of the inner ring agrees very well with the line maps obtained 
with HST by \cite{mao01}, in the sense that most \brg\ hot spots can also
be identified in \ha\ and/or \paa .The only region with significant \hel\ 
flux is located in the inner ring at PA $170\dg$. 
The strongest \htwo\ emission is distributed over the inner $3\as$, with
a hot spot just south of the nucleus.
We do not detect significant flux in the \fetwo\ line within the SINFONI field.

\subsection{IC\,1438} 
At a distance of $33.8\mpc$, the poorly studied Sa spiral IC\,1438 is the most 
distant galaxy in our sample, and with a radius of about $500\pc$, the star
formation ring is at the limit of what we consider ``nuclear''. 

\subsubsection{Continuum emission}
Over the field of view of SINFONI, the K-band contours of IC\,1438 are rather 
elongated outside the very nucleus. This is unlikely to be a projection effect 
because IC\,1438 has a rather low inclination of $i = 32\dg$ \citep{kna06}. 
Indeed, the outer disk appears fairly round in the DSS image. 
On the other hand, the central $10\kpc$ of IC\,1438 are
dominated by a strong stellar bar which is also evident in the I-band
image of \cite{kna06}. It is therefore likely that the gravitational 
potential in the center of IC\,1438 is bar-shaped as well. It appears
that the  PA of the bar within the SINFONI field of view is higher than
that of the large-scale bar. Whether this is the result of isophote
twisting, or whether there is a ``nested'' secondary bar, requires 
additional data.

\subsubsection{Emission lines}
Due to the lower S/N of the SINFONI data, the \brg\ map does not reveal the
nuclear ring as clearly as the \ha\ map, mostly because the fainter, extended
emission is not as obvious. Nevertheless, the two maps agree rather well,
especially in the location of the ``hot spots''. Star formation is strongest 
at the south-east corner (PA $120\dg$) where an extended arc dominates  the
emission in both maps. This region also shows strong \hel\ emission.
Two smaller \brg\ clumps at PA $0\dg$ and $260\dg$ are
also apparent in the \ha\ map. There is also \brg\ emission in the nucleus,
indicating star formation within the central $\approx 50\pc$.

As in the other sample galaxies, the \htwo\ emission is distributed much more
smoothly, and thus outlines the ring structure more completely. There is some
evidence for a connection between the ring and the nuclear region from an
\htwo\ ``ridge'' at PA $260\dg$.

\subsection{Kinematics of Stars and Gas}\label{subsec:kin}
In all five galaxies of our sample, the stellar velocity fields appear rather 
regular, despite the intense star formation within the nuclear rings. 
The PA of the kinematic major axes as listed in Table~\ref{tab:sample} was obtained by 
minimising the difference between the SINFONI velocity fields and a bi(anti)symmetric 
model \citep[for details, see Appendix C in][]{kra06}.
Within the uncertainties, the values are in good agreement with the PA of the 
photometric major axis, at least inside the SINFONI field of view. 
Unfortunately, the limited S/N of our data
does not allow a reliable check for small misalignments and
twists in the velocity fields that might be indicative for the presence of a stellar bar.

Because \brg\ is generally the most prominent emission line in our data, we used
it for an estimate of the ionized gas kinematics. Our analysis of the \brg\ velocity 
fields yields results similar to those derived from the stellar
kinematics. In all cases, velocity gradients within the nuclear rings 
are smooth and the ionized gas has the same sense of rotation as the stars.
Given the limited spatial coverage  of the \brg\ emission, we are unable to
determine whether the ionized gas is kinematically misaligned with respect to
either the stellar rotation field or the photometric major axis. 

In their analysis of NGC\,1079, \cite{mao96} noted that the gas inside the ring rotates
faster than the stars at the same radius. Our observations confirm this result for
all our objects: despite the limited spatial coverage of our \brg\ velocity field, it
is clear from the relevant panels in Figures~\ref{fig:n613} to \ref{fig:i1438}
that the gas generally reaches higher velocities at both ends of the ring. This 
can be understood from the dissipative nature of the gas: because the 
kinetic energy of the gas is conserved, the rotation velocity 
increases while the dispersion decreases. As a consequence, any random motion 
in the z-direction is quickly dampened, and the gas settles in a thin disk.  

As mentioned in \S~\ref{subsec:outflow}, the case of NGC\,613 warrants special 
attention, because the ionized gas is clearly disturbed by the radio jet. 
This is illustrated by the ``twisting'' of the isovelocity contours at the ends 
of the jet, and the increased velocity dispersion of the surrounding gas as seen
most clearly in the dispersion map of the \fetwo\ line presented in Figure~\ref{fig:sigma}.

Two of our targets (NGC\,613 and NGC\,1300) are also present in the 
sample of \cite{bat05}
who derived stellar velocity fields for 23 spirals from integral-field observations 
over a comparable field of view, albeit at optical wavelengths. While their results 
for NGC\,613 agree well with ours, both in shape and amplitude of the stellar kinematics,
there are significant differences for NGC\,1300. More specifically, the stellar
rotation field of \cite{bat05} (their Figure~3) appears ``flipped'' by $180\dg$ with respect
to ours, i.e. it rotates in the opposite sense. We also note that their optical continuum map 
seems rotated by $90\dg$. We believe that the maps presented in this study have the correct 
orientation, based on the comparison with the \ha\ map of \cite{kna06} which agrees very well 
with our \brg\ map. 

\subsection{Near-Infrared spectra of the nuclei}\label{subsec:nuclei}
In general, the nuclei (i.e. the central $1\as$) of our sample galaxies appear
rather quiescent. Except for NGC\,613, which we will discuss in more detail
below, the NIR spectra of the nuclei in Figures~\ref{fig:n613} to \ref{fig:i1438} 
show little, if any, evidence for NIR emission lines. There is 
faint nuclear \fetwo\ and \htwo\ emission apparent in the line maps of NGC\,1300
(Figure~\ref{fig:n1300}), as well as some \brg\ emission in IC\,1438 nucleus
(Figure~\ref{fig:i1438}). However, since these emission lines are not detected in
the respective spectra, the apparent emission is likely a residual due to 
imperfect continuum subtraction. In Table~\ref{tab:flux}, we therefore list 
only upper limits for the emission line fluxes from the galaxy nuclei (except for
\fetwo\ and \htwo\ in NGC\,613).

The lack of gaseous emission lines is not unexpected in a scenario in which
gas accumulates at the nucleus over time until a critical density is reached. 
Star formation is then triggered, but can continue only until the gas supply 
is consumed or the energetic outflow from supernovae explosions disrupts 
and/or expels the gas, thus ``quenching'' star formation. 
This scenario is entirely consistent with observations, at least for late-type
spirals: only about  10\% of nuclear star clusters currently show emission
lines, although most of them harbor a young stellar population less than
100\,Myrs old  \citep{wal06,ros06}. This can only be  explained if star
formation in galactic nuclei is episodic in nature, with a duty cycle of about
10\%.

This scenario appears to be supported by the case of
NGC\,613. The NIR spectrum of its nucleus shows several strong emission lines
from the \htwo\ molecule, as well as strong emission from \fetwo\  at $1.644\mum$. 
However, the \brg\ recombination line is not detected at our sensitivity limit,
implying that star formation does not currently take place. 
This implies that while molecular gas is present in the nucleus, it has not 
sufficiently cooled to reach the critical density for star formation. 
The most likely excitation mechanism for both \htwo\ and \fetwo\ in the nucleus
of NGC\,613 is the mechanical energy delivered by the AGN outflow.
In the case of \fetwo , this is different from the hot spots 
in the ring, where it is more likely produced by supernova remnants.
In paper II, we will conduct a more detailed analysis of the physical state 
of the emission line gas in the nucleus of NGC\,613.

\section{Discussion: how does star formation proceed along a ring?}\label{sec:discuss}
\begin{figure}
\begin{center}
\includegraphics[angle=270,scale=.8]{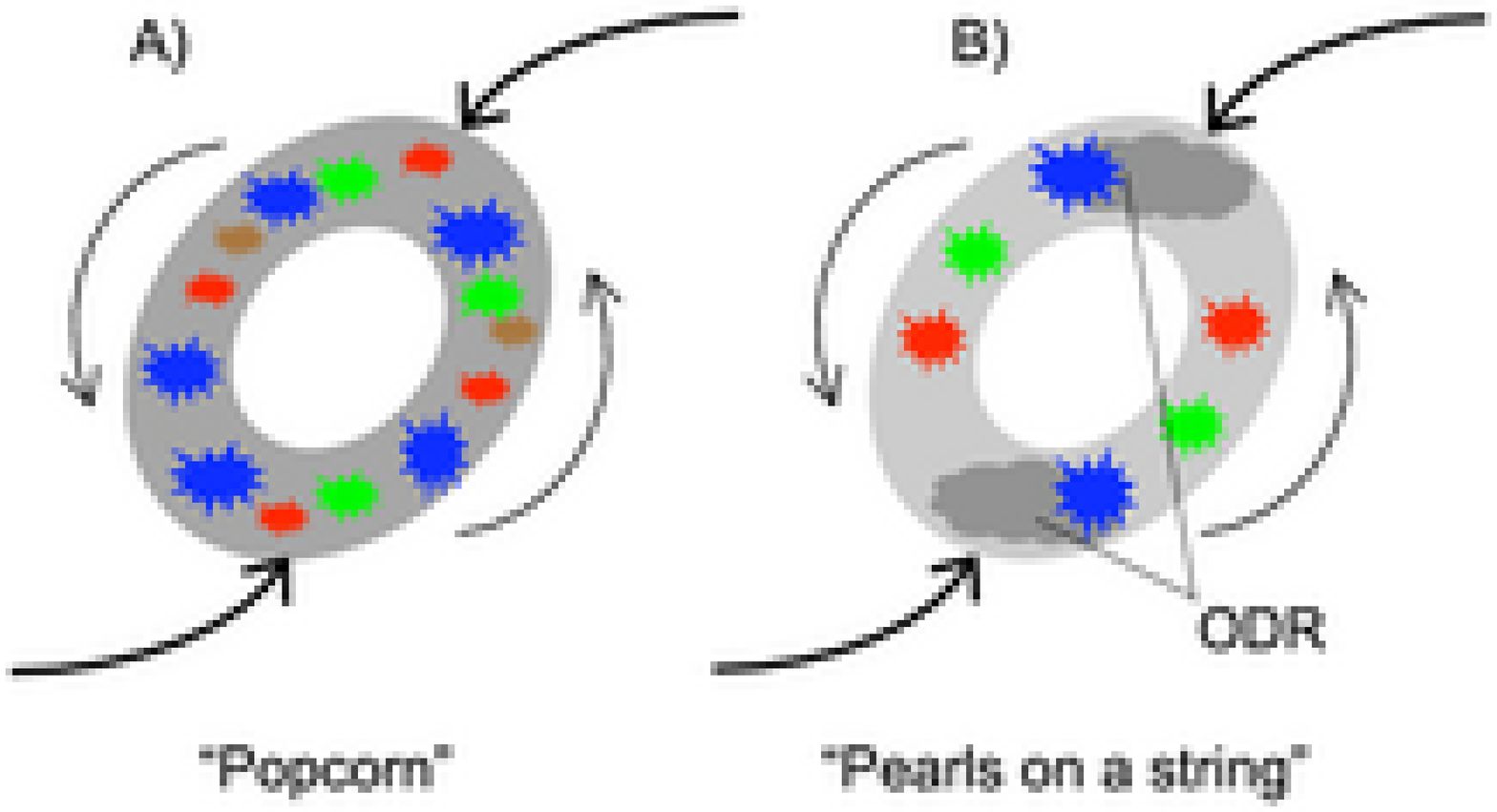}
\caption{Illustration of two candidate scenarios for star formation
in a nuclear ring. Dark grey areas denote dense, cold gas that
is conducive to star formation. The various star symbols denote young
stellar clusters, their colors signify the cluster age in the sense
that age increases from blue to green to red. A clear age sequence is
expected only in the ``pearls on a string'' scenario 
(see \S~\ref{subsec:pop} for discussion).}
\label{fig:pop}
\end{center}
\end{figure}

The basic picture of star formation rings being the consequence of gas
responding to a bar-shaped potential is generally accepted, and is
nicely illustrated e.g. in \cite{ath92}: dynamical resonances
associated with the non-axisymmetric rotating bar potential cause the
orbiting gas to lose angular momentum and thus ``spiral'' towards the
nucleus. In the process, the gas is concentrated in two elongated and
curved dust lanes along the leading sides of the bar. In optical
images, it appears as if the dust lanes connect the end of the bar to
the nuclear ring. When observed at high angular resolution, the ``ring'' 
actually more closely resembles two tightly wound spiral arms. 
At the ``contact points'' between the dust lanes and the ring, the gas becomes 
less turbulent, and enters the almost circular orbits delineating the
ring. While it is clear that there is abundant (molecular) gas throughout
the ring, there is some debate about how and where star formation occurs. 
At least two mechanisms are plausible as illustrated in Figure~\ref{fig:pop}, 
which we briefly describe in the following.

\subsection{``Popcorn'' or ``Pearls on a String''?}\label{subsec:pop}
In the first scenario, the gas accumulates along the ring, until a critical
density is reached.  Presumably driven by turbulence \citep{elm94}, the ring
then becomes unstable to gravitational collapse, and star formation is induced.
In this model, either the entire ring structure begins to form stars at the
same time, or individual hot spots collapse at random times and locations
within the ring. Either way, there should be no systematic age sequence of hot
spots along the ring. Because of its stochastic nature, we call this the
``popcorn model'' of star formation. 

In the second scenario, a short-lived, quasi-instantaneous ``delta burst'' of
star formation is induced in (and only in) a well-defined region
along the ring where the gas density becomes sufficiently high to ignite
star formation. The precise location of this ``overdensity region'' (ODR)
depends on the details of the potential, but it is often found close
to, and somewhat downstream from where the gas enters the ring, as 
illustrated e.g. in Figure~15 of \cite{reg03}. 

A young cluster that has formed at this location will continue
its orbit along the ring, but star formation will cease as soon as the first
supernova explosions expel the gas, i.e. after a few million years. 
Because the gas entering the ring is intrinsically clumpy, a series of 
short-lived starbursts will be triggered at the ODR. After star formation
ends, the cluster will age passively as it follows its orbit. 
A series of such events will then produce a sequence of star clusters 
that enter the ring  like ``pearls on a string''. In
this scenario, the star clusters should show a bi-polar age gradient around
the ring, with the youngest clusters found close to the ODR, and
increasingly older cluster ages in the direction of rotation, up to the
opposite ODR. Because the age differences are simply due to the
travel time along the ring, they should be consistent with the observed
velocities of gas and stars in the ring, a question we briefly discuss in
\ref{subsec:pearls}.

Evidence for such age gradients has been found in a few cases. The most 
convincing case is the nuclear ring in M\,100, for which \cite{ryd01} and
\cite{all06} used NIR and optical absorption indices and emission lines to
determine the stellar ages of the hot spots in the star formation ring. 
\cite{smi99} have claimed evidence for sequential star formation in the 
nuclear ring of NGC\,7771 from the ratio of NIR to radio continuum. 
Moreover, \cite{dav98} have estimated 
the ages of individual hot spots in the ring of NGC\,1068, based on the 
equivalent width of the \brg\ emission. Although these authors do not
specifically comment on this, their Table\,2 shows hints for a bi-polar age
gradient also in this star formation ring, although the age differences appear
too small to be significant. Finally, \cite{maz07} find evidence for such age
gradients in 10 star formation rings out of a sample of 21. 

\subsection{A new method for estimating hot spot ages}\label{subsec:ages}
In general, measuring the stellar cluster ages along the ring is difficult,
even in a  relative sense. The studies mentioned above have relied either on
stellar absorption indices or equivalent width (EW) measurements of hydrogen
recombination lines. The advantage of both diagnostics is that they are
insensitive to extinction (but not to continuum emission from hot dust).
According to the STARBURST99 models \citep{lei99}, the equivalent width of 
hydrogen recombination lines produced in an instantaneous burst 
is more or less constant until about 3\,Myrs after the burst, and then
drops steeply with time. It falls below the detection limit (about 1\,\AA ) 
of typical observations after only $\approx$10\,Myrs. 
Using e.g. the EW of \ha\ is therefore limited to clusters in the age range
$3 - 10\myr$, and any inferred age differences are small (few Myrs) by definition.

Moreover, there is a fundamental difficulty with this kind of analysis: it
requires accurate knowledge of the continuum emission from the hot spot itself,
i.e. after subtraction of the underlying bulge and/or disk.
In the centers of spiral galaxies, this separation between hot spot and bulge/disk
is extremely difficult. The continuum emission from the bulge/disk is at least 
as bright as that from the hot spot itself and often much brighter, especially 
at NIR wavelengths where the light from evolved stars dominates. 
Because of the small field of view of integral field units, it is normally 
not possible from the data itself to disentangle the various components. 
Accurate continuum subtraction therefore often has to rely on uncertain
assumptions on the galaxy structure. While the intrinsic symmetry of the 
bulge/disk will reduce the error for relative age measurements, 
estimating {\it absolute} ages of young clusters in circumnuclear star formation rings 
using absorption indices or emission line EWs is clearly fraught with uncertainty.

Here, we use an alternative approach to obtain at least a qualitative sense for
whether or not the young clusters in the ring show systematic relative age gradients. 
Instead of EWs, we use the flux of three emission lines that are prominent in 
the NIR spectra of star forming regions: \hel , \brg , and \fetwo . Because line
fluxes are independent of the underlying continuum level, they can be measured
accurately\footnote{None of the emission lines discussed here is significantly 
affected by possible stellar absorption features ``underneath''
the line. In fact, we did subtract the best-fitting mix of stellar
templates before measuring the line fluxes listed in Table~\ref{tab:flux}.
However, this made little or no difference to obtaining fluxes from the 
spectra directly (i.e. without prior subtraction of the stellar continuum).}
without any knowledge of the disk/bulge structure.  
The assumption here is that the emission lines are produced only in the 
ring itself, i.e. that the underlying bulge/disk can be considered quiescent. 
Given that the line maps in Figs.~\ref{fig:n613} to
\ref{fig:i1438} show very  little, if any, emission outside the ring proper,
this seems a rather safe assumption for our targets. 

As mentioned in \S~\ref{sec:diagnostics}, the \brg\ and \hel\ lines are both
produced by photo-ionization in the Str\"omgren-spheres around O- or B-type
stars. Because the \hel\ line has a higher ionization energy than \brg , it
requires the presence of hotter and more massive stars, and hence its flux
falls off more rapidly after an instantaneous burst of star formation than
that of \brg . The ratio of \hel\ to \brg\ flux can thus be used as a
qualitative age indicator for cluster ages, in the sense that younger clusters
still harbor hotter stars, and thus show more \hel\ emission. We emphasize
that that this is a qualitative statement only, and that we make no attempt to
use the \hel /\brg\ ratio to derive absolute ages, because the detailed
interpretation of this line ratio is complicated \citep{doh95,lum01}.

Assuming that all clusters have a similar initial mass function (IMF), and are
sufficiently massive for statistical variations in stellar inventory to be
insignificant, this diagnostic is useful for cluster ages up to about $10\myr$.
Around that time, the supernova rate for an instantaneous burst reaches its
maximum \citep{lei99}. As described in \S~\ref{sec:diagnostics}, the fast
shock fronts associated with these supernovae are efficient in creating  the
extended partially ionized zones that produce strong \fetwo\ emission. Relative
to \hel\ and \brg\ (which both have subsided significantly by this time), the
\fetwo\ line becomes dominant for ages above $\approx 10\myr$. 

The range of cluster ages probed by this approach is $0 - 35\myr$, 
when the supernova rate drops almost instantaneously \citep{lei99}. 
This time scale is significantly larger than that probed by EW measurements,
and is well matched to the expected travel
time of gas and star clusters around the ring: with a rotation velocity  of
$100\kms$, it would take about $15\myr$ to complete a half orbit for a ring
diameter of $1\kpc$. As can be seen from Table~\ref{tab:rings}, most of the
rings in our  sample are even smaller than this. The typical rotation
velocities of the gas (which  better probes the ring material) are of order
$100\kms$ (see the velocity maps of \brg\ or the stars in Figures\,\ref{fig:n613} 
to \ref{fig:i1438}, and so it is entirely plausible that any age differences 
probed by these emission lines are indeed reflecting the travel
times of clusters on their orbit along the ring.

Finally, we note that for these relatively short orbital timescales, the ring 
diameter can be regarded as constant. While \cite{reg03} have shown that nuclear
rings become smaller over time due to dissipation of angular momentum,
this occurs over timescales of hundreds of Myrs, and thus can be ignored for
the the present discussion.

\subsection{Application to the data}\label{subsec:pearls}
In order to look for evidence of systematic age gradients as predicted by 
the ``pearls on a string'' scenario described in \S~\ref{subsec:pop}, we 
first need to identify the location of any ODRs, if present. 
This is not a straightforward task, because maps and velocity fields of the 
molecular gas are difficult to obtain, especially for sources in the southern 
hemisphere, and reach the required resolution only in a few nearby galaxies 
\cite[e.g.][]{sch03,sch06}. Lacking molecular gas observations, 
the best approach is to use high spatial resolution 
maps of the dust morphology, as demonstrated by \cite{mar03}.
We have therefore searched the HST archive for optical images.
Whenever such imagery was available, we identified dust lanes,
their intersection with the ring, and any region(s) of high extinction
along the ring that could plausibly be identified with an ODR. 
In cases where only one ODR was obvious, we have assumed that the 
second is located on the opposite side of the ring.

\begin{figure}
\begin{center}
\plottwo{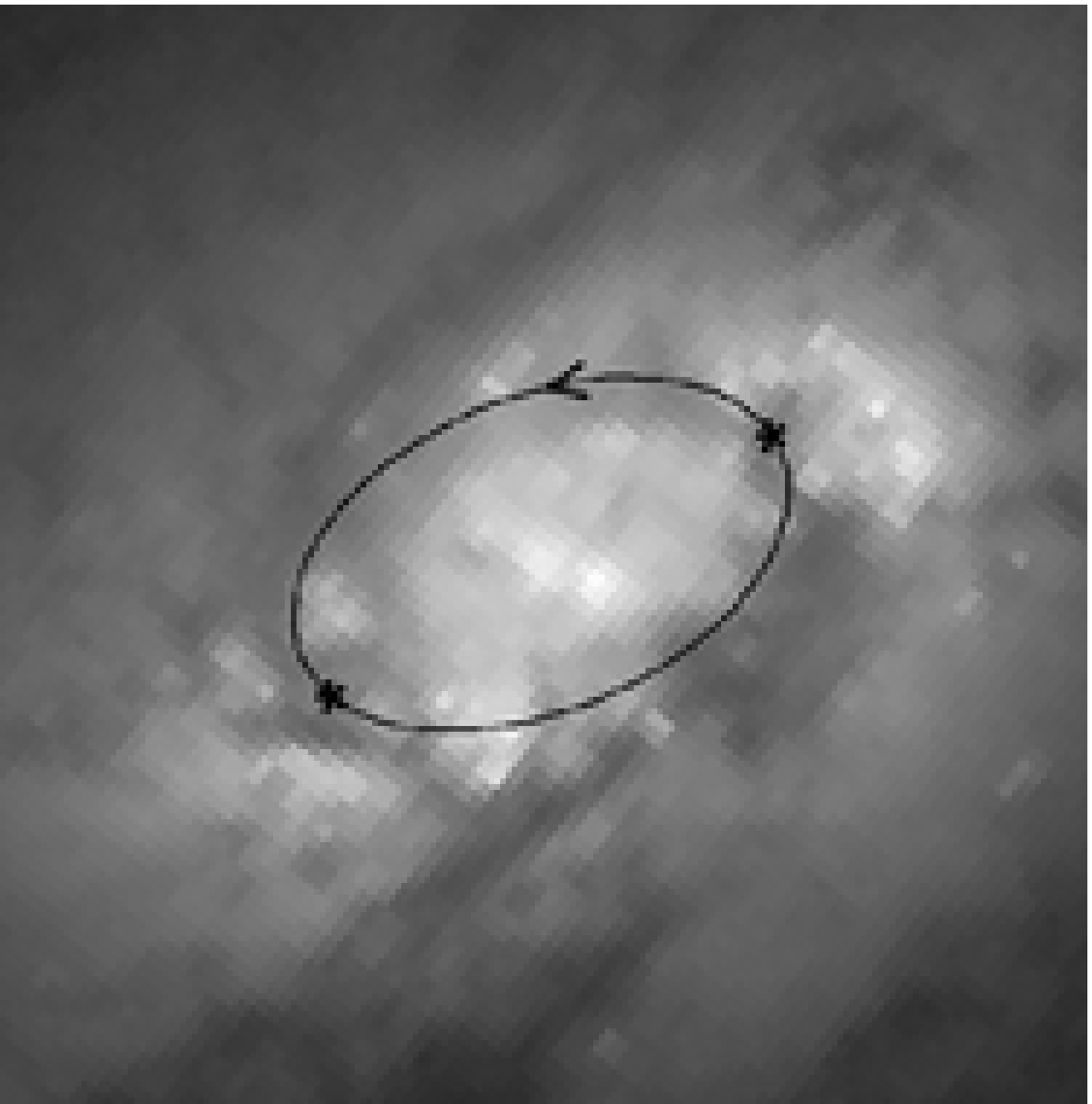}{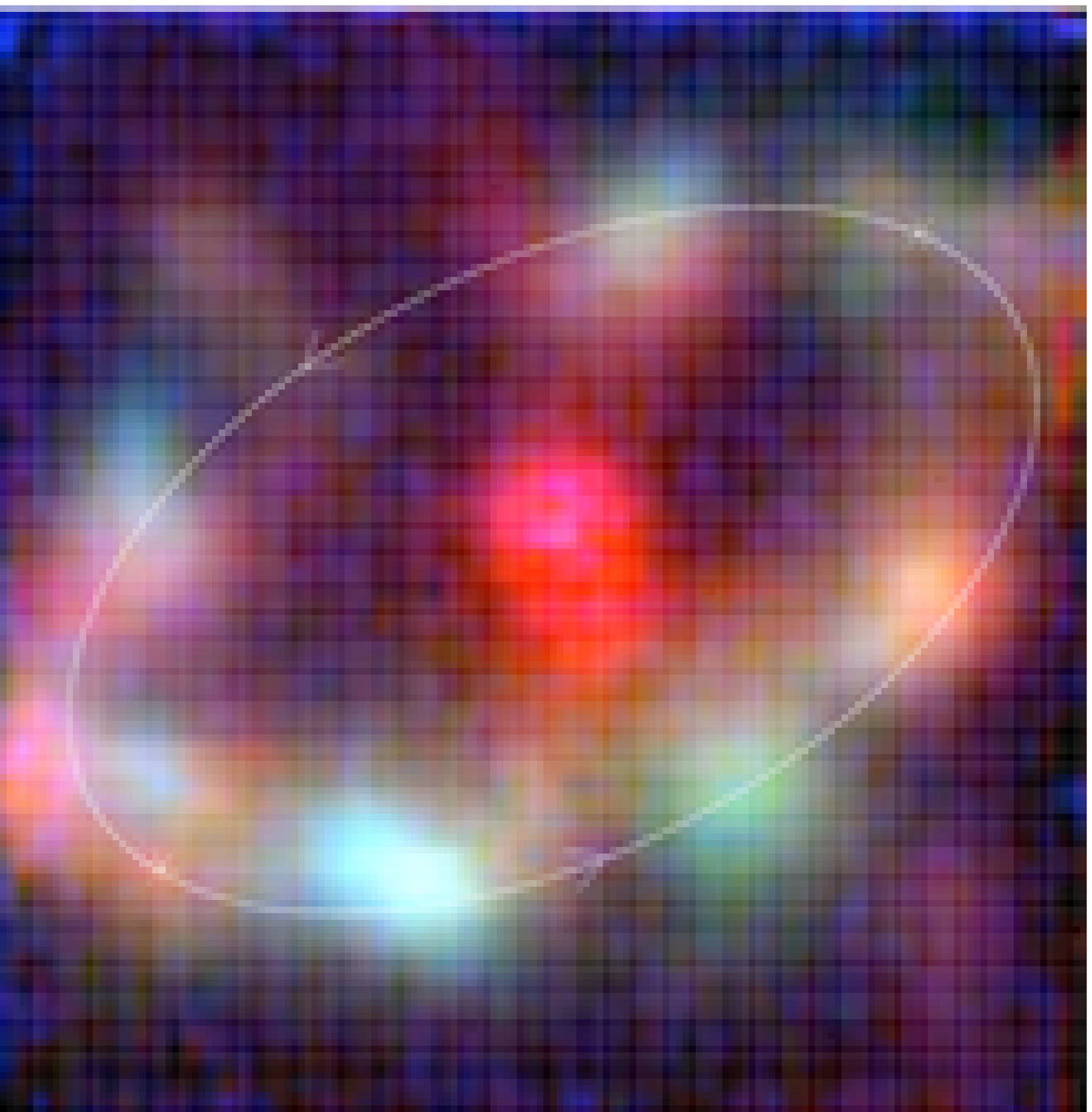}
\caption{Left: Archival HST/WFPC2 F606W map of the inner $15\as \times 15\as$ of 
NGC\,613. An ellipse with the parameters listed in Table~\ref{tab:rings} 
is overplotted in order to outline the assumed flow path of
matter in the star formation ring. Also indicated are the assumed
locations of overdensity regions (ODRs, marked by star symbols) where 
infalling gas is compressed. The sense of rotation is indicated with an arrow. 
Right: false color map of the emission line morphology in the inner 
$8\as \times 8\as$ of NGC\,613. Here, the blue channel represents 
\hel , green \brg , and red \fetwo\ emission. Note that - at least 
in the southern half of the ring - the hottest stars (as marked by \hel\ 
emission in blue) are located immediately downstream from the contact 
points. In contrast, the supernovae (as
traced by the \fetwo\ emission in red) occur predominantly further along the
ring, with \brg\ emission dominated the region in between. This suggests that
star formation is triggered at the ODRs, and that the star forming
clumps then age passively along their orbit.}
\label{fig:n0613_color}
\end{center}
\end{figure}

Figure~\ref{fig:n0613_color} shows the case of NGC\,613. The left panel contains
the HST/WFPC2 V-band (F606W) image of the central $15\as (1.3\kpc)$. 
Two prominent dust lanes are easily identified, and the inferred approximate
locations of the ODRs are marked by star symbols. Note that the small size of the
symbols certainly does not imply a positional accuracy, but merely
indicates a rough estimate. In fact, the ODRs can be quite extended as 
demonstrated in the simulations of \cite{reg03}.

In both panels, the star formation ring is indicated by an ellipse 
with the ring parameters as listed in Table~\ref{tab:rings}. Here, and for all other
galaxies, the sense of rotation was determined by assuming that the outer spiral arms
are trailing. In the right panel, we present a false-color (RGB) image constructed from 
the SINFONI emission line maps of \hel , \brg , and \fetwo , assigned to the blue, 
green, and red color channel, respectively. 

At least for the southern half of the ring, the color map appears to show the exact 
trend predicted by the ``pearls on a string'' scenario: starting at the ODRs,
and moving in the direction of the gas flow,  
the color sequence is blue-green-red. As discussed in \S~\ref{subsec:ages}, this is
exactly as expected if the clusters age passively along their orbits. The trend is
less pronounced in the northern half of the ring, but here the picture is complicated
by the strong mechanical interaction between the jet and the ring. It is possible that
any cluster located within the ``gap'' at P.A. $30\dg$ has been stripped of its gas, and
thus does not show up in the line maps, even though it might be relatively young.

\begin{figure}
\begin{center}
\plottwo{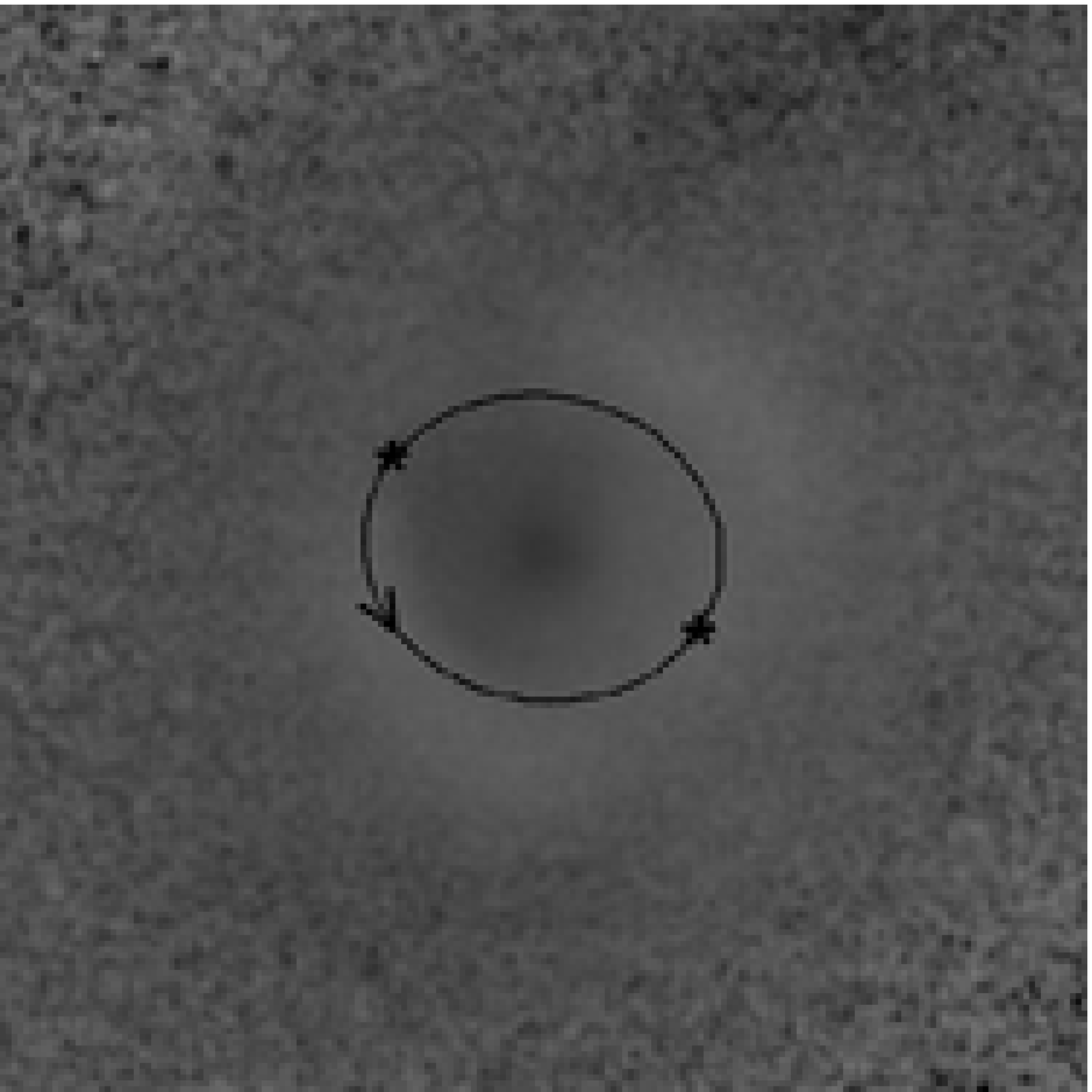}{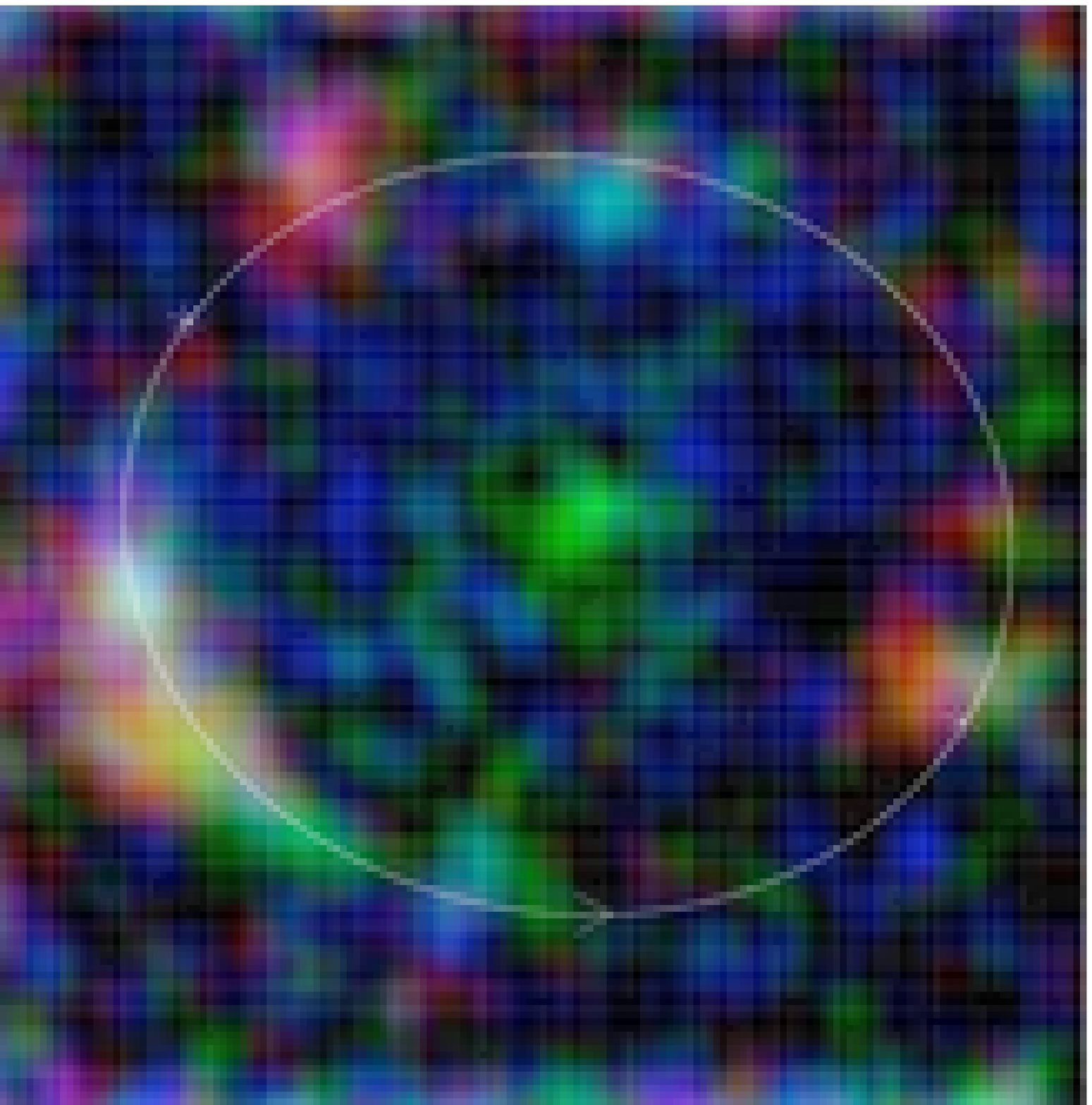}
\caption{Left: relative (i.e. uncalibrated) B-I color map of the inner 
$20\as \times 20\as$ of IC\,1438. An ellipse with the ring parameters as
listed in Table~\ref{tab:rings} is overplotted in order to outline the 
assumed flow path of
matter in the star formation ring. Also indicated are the assumed
locations of ODRs (star symbols) where infalling gas is compressed. 
The sense of rotation is indicated with an arrow. 
Right: false color map of the emission line morphology in the inner 
$8\as \times 8\as$ of IC\,1438 as described in Figure~\ref{fig:n0613_color}. 
}
\label{fig:i1438_color}
\end{center}
\end{figure}

For a variety of reasons, the corresponding analyis for the other sample galaxies
yields a somewhat less convincing picture. The case of IC\,1438, presented in 
Figure~\ref{fig:i1438_color}, suffers from the fact that is has never been imaged
with HST. In order to gain insight into the dust morphology, we therefore used the 
B- and I-band ground-based images presented in \cite{kna06}
to create a B-I ``pseudo'' color map\footnote{The term ``pseudo'' is used here because 
no reliable magnitude scale can be attributed to the color map presented in 
Figure~\ref{fig:i1438_color}. The reason is that
a careful sky subtraction was not possible for the \cite{kna06} images because 
most galaxies completely filled the field of view of the CCD used. Nevertheless, 
the map is still valid in a relative sense, and can be used to identify ``redder'' 
regions that likely have higher extinction.}. While the ring stands out as very 
blue in B-I (i.e. bright in Figure~\ref{fig:i1438_color}), 
outside the ring the dust morphology is rather smooth. 
The only discernible structure is a faint region of higher extinction 
north of the nucleus which we interpret as the arm of a gas spiral. 
We also assume that one ODR is located close
to the intersection of this spiral and the ring, and that the other lies opposite 
to it. With these - admittedly rather uncertain - assumptions, however, the
picture for IC\,1438 is similar to that for NGC\,613: \hel\ peaks close to
the eastern ODR, while \fetwo\ is strong in clusters that have traveled a good
distance around the ring.

\begin{figure}
\begin{center}
\plottwo{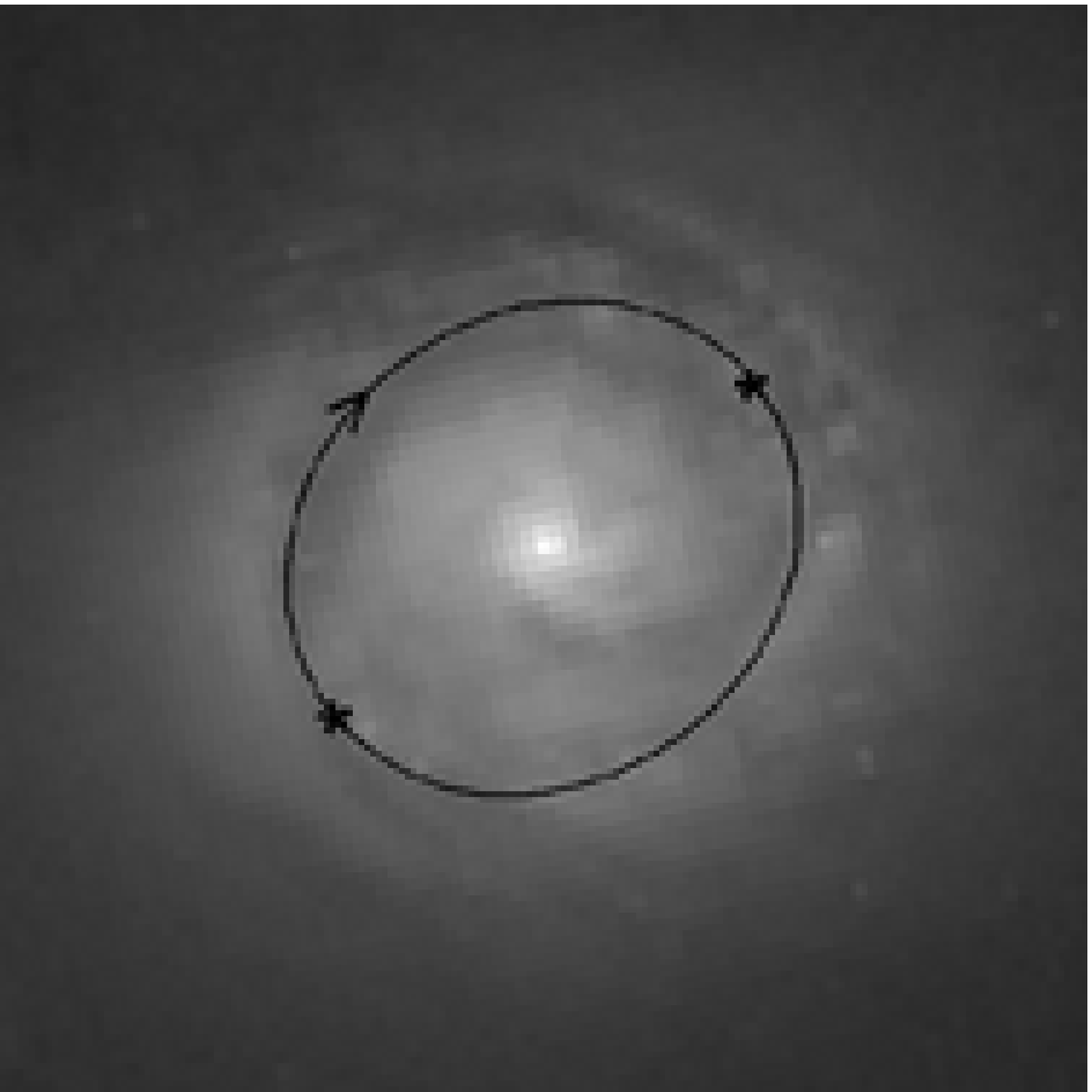}{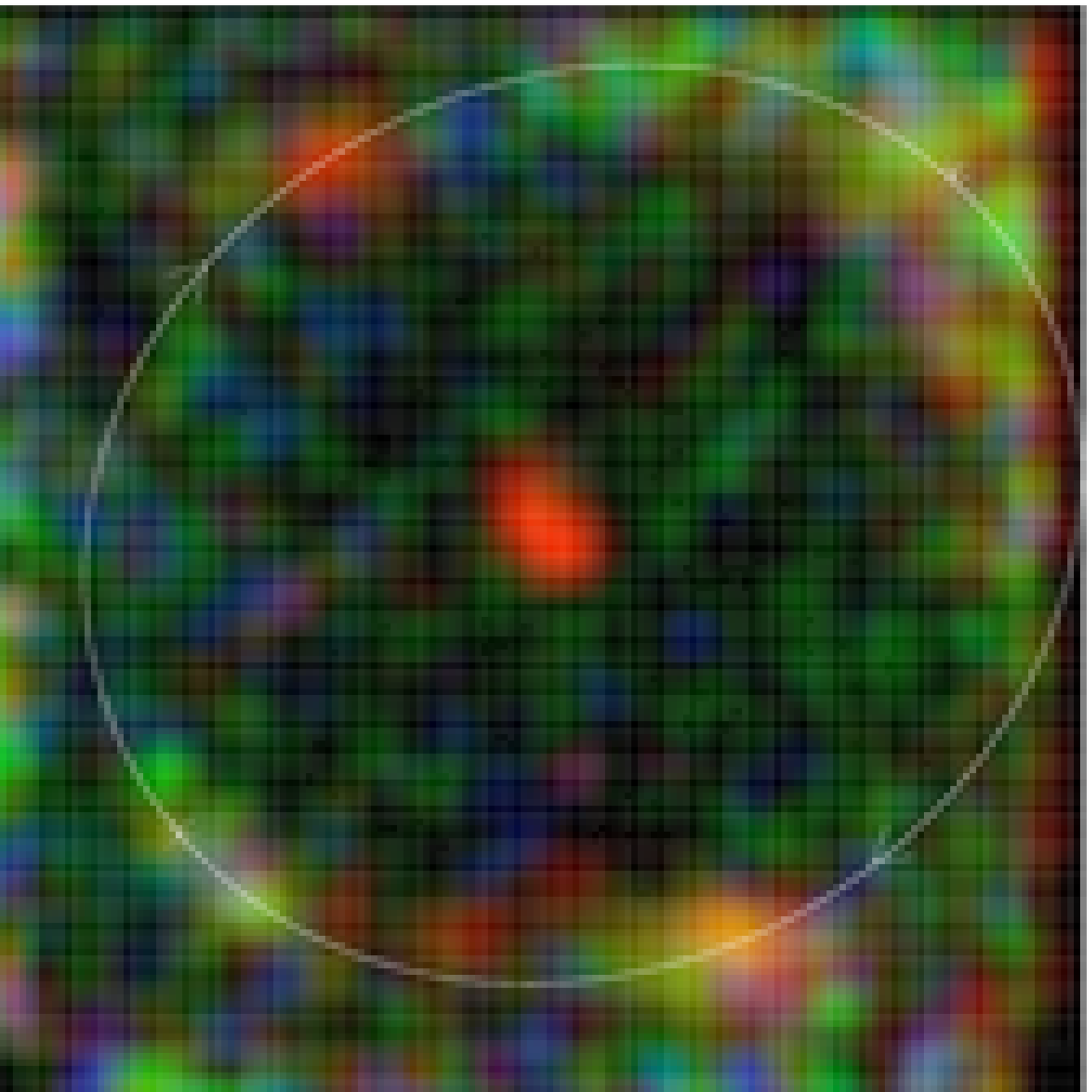}
\caption{Left: Archival HST/WFPC2 F606W map of the inner $15\as \times 15\as$ of 
NGC\,1300. An ellipse with the parameters listed in Table~\ref{tab:rings} 
is overplotted in order to outline the assumed flow path of
matter in the star formation ring. Also indicated are the assumed
locations of ODRs (star symbols) where infalling gas is compressed. 
The sense of rotation is indicated with an arrow. 
Right: false color map of the emission line morphology in the inner 
$8\as \times 8\as$ of NGC\,1300 as described in Figure~\ref{fig:n0613_color}. 
}
\label{fig:n1300_color}
\end{center}
\end{figure}

The case of NGC\,1300 is presented in Figure~\ref{fig:n1300_color}. Because only one 
OB was executed for this object, the quality of the SINFONI line maps is less 
than optimal. This is particularly true for the \hel\ map which only shows tentative 
evidence for two emission peaks at P.A. $135\dg$ and $315\dg$. Intriguingly, however, 
this is exactly where one would place the ODRs based on the dust morphology derived 
from the HST/WFPC2 F606W map.
Again, \fetwo\ emission peaks are located far from the assumed ODRs.

\begin{figure}
\begin{center}
\includegraphics[angle=0,scale=.65]{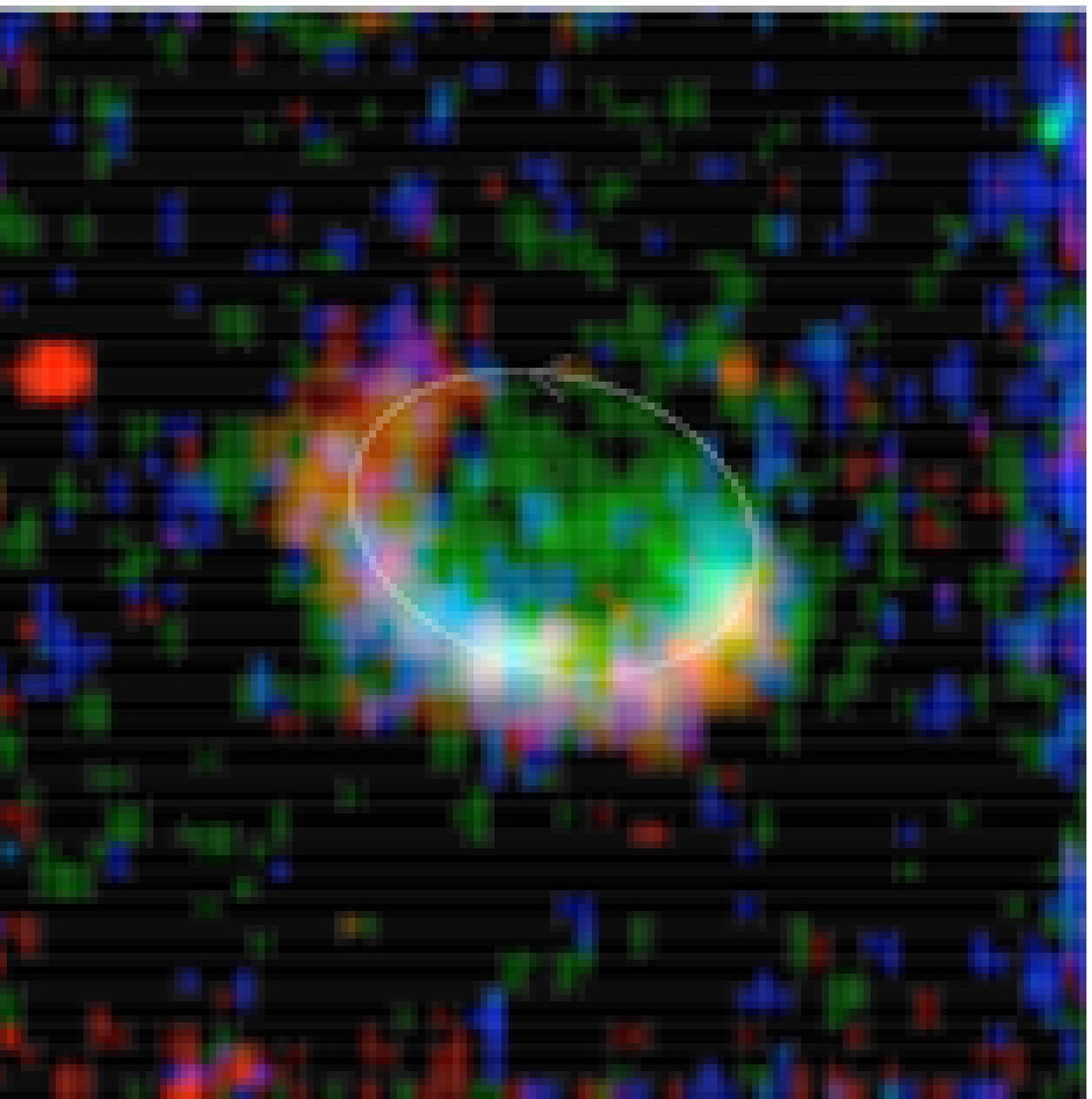}
\caption{False color map of the emission line morphology in the inner 
$8\as \times 8\as$ of NGC\,1079 as described in Figure~\ref{fig:n0613_color}. 
The sense of rotation is indicated with an arrow. 
}
\label{fig:n1079_color}
\end{center}
\end{figure}

Figure~\ref{fig:n1079_color} shows the false color map of the emission line morphology
in the ring of NGC\,1079. For this galaxy, there is neither HST imagery available, nor was
it possible to identify any structure from a ``pseudo'' color map as described for IC\,1438.
We therefore are unable to estimate the location of the ODRs, and therefore cannot 
use this galaxy to test for a systematic age gradient along the ring.

\begin{figure}
\begin{center}
\includegraphics[angle=0,scale=.65]{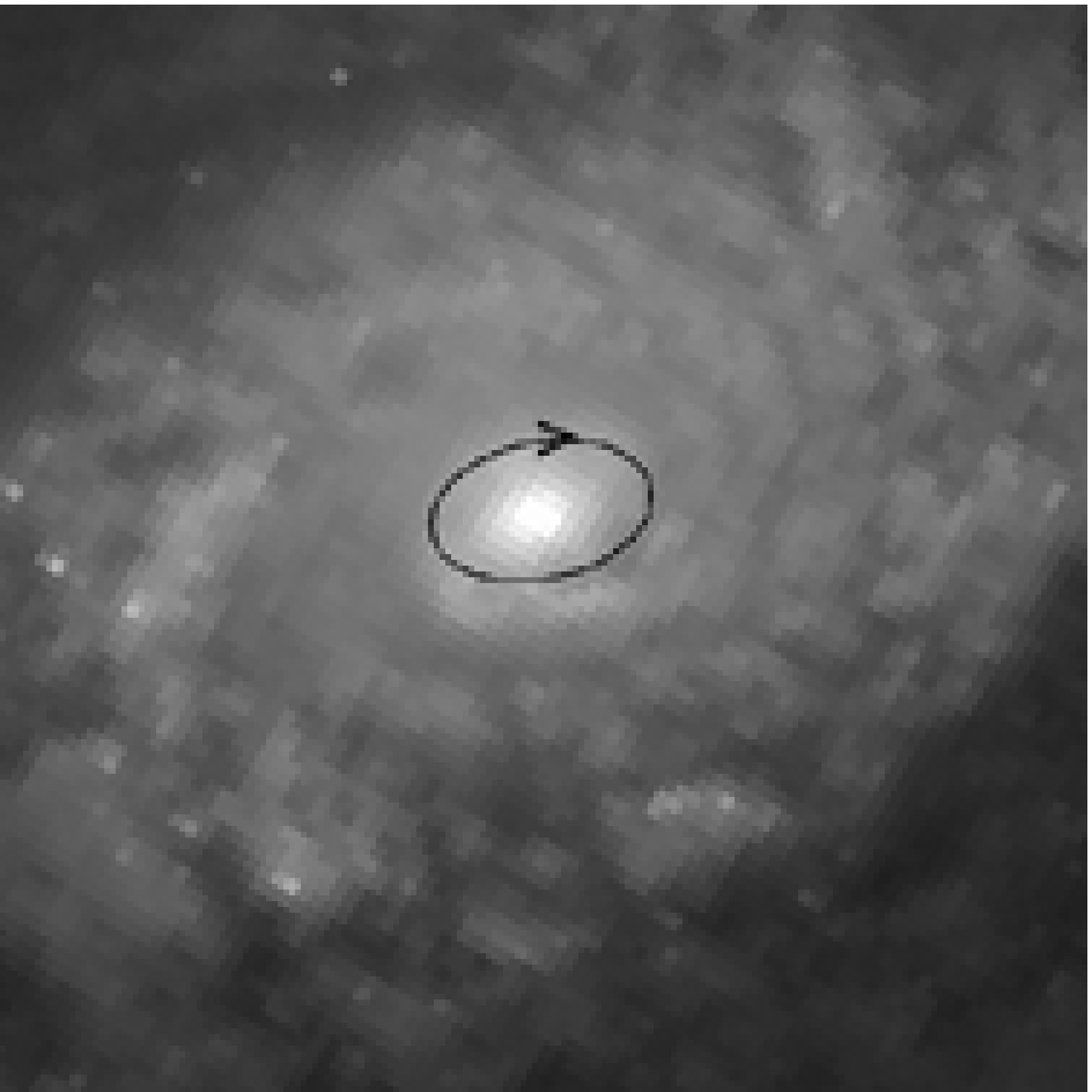}
\caption{Archival HST/WFPC2 F814W map of the inner $15\as \times 15\as$ of 
NGC\,5248. An ellipse with the parameters listed in Table~\ref{tab:rings} 
is overplotted in order to outline the assumed flow path of matter in the 
inner star formation ring. The sense of rotation is indicated with an arrow. 
We were unable to even tentatively identify distinct ODRs in this object.
}
\label{fig:n5248_hst}
\end{center}
\end{figure}

Finally, the case of NGC\,5248 does not appear to be a valid test of the 
proposed scenario because it has two nested rings, 
the larger one of which falls mostly outside the SINFONI field of view (but can be recognized 
in the line maps of Figure~\ref{fig:n5248}). Because any gas falling onto the inner ring must first
pass through the outer ring, and will be disturbed by star formation in the
outer ring, the gas dynamics in NGC\,5248 are likely very different from those
in the other galaxies. This is possibly the reason why the dust morphology as shown 
in Figure~\ref{fig:n5248_hst} also appears different from that
in the cases dicussed above. NGC\,5248 harbors a ``flocculent'' nuclear dust spiral
without any clearly defined dust lanes \citep{lai01}. 
Because of the lack of dominant dust features, 
we were unable to determine the location of any putative ODRs. 

In summary, three of the five galaxies in our sample show line ratios that
appear consistent with the ``pearls on a string'' scenario of how star formation 
progresses in nuclear rings. The data for the remaining two galaxies are incomplete,
but not inconsistent with either model. 

\subsection{Comparison to other studies}\label{subsec:synergy}
Our findings lend some support to a picture in which young stars in nuclear
starburst rings predominantly form in two distinct regions
close to the ``contact points'' between the dust lanes and the ring. 
This is consistent with observations of the molecular
gas distribution in the nuclei of barred galaxies which often 
shows a ``twin peaks'' geometry \citep[e.g.][]{ken92}. 
This implies that cold molecular
gas (i.e. the star formation fuel) exists only in two locations along the
ring. Other studies claiming evidence for an age sequence of hot spots
in nuclear rings have been mentioned in \S~\ref{subsec:pop}. On the other hand,
\cite{mao96} have found that the starburst rings in NGC\,1079 and NGC\,5248
experienced star formation over several $10^8\yr$. 
This apparent contradiction might be explained by taking into account
the dynamical timescales. As discussed
in section \ref{subsec:pop}, it takes as little as $15\myr$ for a stellar
cluster in the ring to rotate around half the ring. Therefore, 
several generations of stellar clusters will mix and overlap 
after typically $30-50\myr$. These older
clusters are still prominent in broad band images (but not in
the emission line maps discussed here). Using color diagnostics 
therefore might lead to a picture of ``random'' star formation in
the ring, even if the ``pearls on a string'' scenario is correct. 
This highlights the importance of using age diagnostics
that are well-matched to the orbital time scales.

\section{CONCLUSIONS\label{sec:summary}}
Based on high-resolution NIR integral-field observations of five 
nuclear star formation rings, we have presented their emission line
morphologies and velocity structure both in gas and stars.
We have introduced a new method to derive relative hot spot ages
along the rings using the relative strengths of the \hel , \brg , 
and \fetwo\ lines. We employ this method to investigate the plausibilty
of two competing scenarios for the way star formation is induced
in nuclear rings, namely i) the ``popcorn'' model in which hot spots
appear stochastically around the ring, and ii) the ``pearls on a string'' 
scenario in which star formation is triggered predominantly at two overdensity 
regions on either side of the ring. Only the latter predicts a well-ordered age
sequence of hot spots along either half of a ring.

The data presented in this study provide tentative support for the
``pearls on a  string'' scenario, in that three out of five sample galaxies
show some evidence for an age gradient of hot spots along the ring, while the
remaining two galaxies have incomplete information and thus are consistent
with either model. 

Of course, it might well be that star formation proceeds differently
in some rings than in others. However, given that nuclear rings
appear to form via a common mechanism (i.e. the gas response to a
bar-shaped potential), it is not unreasonable to expect that they
also follow a common path for inducing star formation.
The small number of objects described here is clearly 
insufficient to provide reliable statistics, and similar studies of larger 
galaxy samples are needed to decide whether a particular mechanism 
governs the star formation in the majority of nuclear rings.
Be that as it may, the proposed method of using NIR line ratios to estimate
relative ages of young star clusters has been demonstrated to be a 
powerful tool for studies along this line.
\acknowledgments 
We gratefully acknowledge the contributions of Emma Allard to the early analysis 
of the NGC\,0613 data. We especially wish to thank A. Lan{\c c}on who kindly 
provided the library of synthetic spectra prior to publication, and resampled 
the model spectra to the resolution of the SINFONI data. 
We are also grateful to R. Davies for providing his code for OH line subtraction,
and to the ESO staff at Paranal Observatory who conducted the observations.
This project made use of the HyperLeda and NED databases. GANDALF was developed
by the SAURON team and is available from the SAURON website
(www.strw.leidenuniv.nl/sauron). The Digitized Sky Surveys were produced at the
Space Telescope Science Institute under U.S. Government grant NAG W-2166. The
images of these surveys are based on photographic data obtained using the
Oschin Schmidt Telescope on Palomar Mountain and the UK Schmidt Telescope.

\begin{table*}
\centering
\begin{tabular}{lcccccccccc}

\hline
(1)  &(2) 	& (3) 	   & (4)   & (5)        & (6) 	 & (7) 	    & (8)     & (9)	   & (10)       & (11) \\
Galaxy  & Type 	& Activity & $D$   & scale      & $M_B$  & $i$ 	    & PA      & PA$_{kin}$ & PA$_{bar}$ & $l_{bar}$ \\
     & 	        &  	   & (Mpc) & (pc/$\as$) & (mags) & ($\dg$)  & ($\dg$) & ($\dg$)    & ($\dg$)    & ($\as$)     \\
\hline
NGC\,613  & SB(rs)bc  & Sy      & 17.5  &  84.8 & -20.53  & 41 & 120 & 120 & 127$^1$ & $> 59^1$  \\
NGC\,1079 & SAB(r'l)a &         & 16.9  &  81.9 & -18.83  & 52 &  87 &  65 & 122$^1$ &  32$^1$	\\
NGC\,1300 & SB(s)bc   &         & 18.8  &  91.1 & -20.42  & 49 & 106 & 116 & 100$^2$ &  94$^2$	\\
NGC\,5248 & SB(rs)bc  & Sy2 HII & 22.7  & 110.1 & -21.07  & 44 & 110 & 105 & 105$^3$ &  95$^3$	\\
IC\,1438  & SAB(r)a   &         & 33.8  & 163.9 & -20.08  & 32 & 145 & 158 &  n.a.  &  n.a.	\\
\hline
\end{tabular}
\caption{
Global parameters of the five sample galaxies, obtained from the RC3 \citep{dev91}
unless otherwise indicated. Tabulated are the galaxy name (col.~1); the morphological
type (from NED; col.~2); nuclear activity class (NED; col.~3); distance $D$ in
Mpc from Nearby Galaxies Catalog \citep[][col.~4]{tul88}; image scale in parsec
per arcsec, as derived from the distance (col.~5); absolute blue magnitude
\citep[][col.~6]{tul88}; inclination $i$ as derived from the ratio of the major
to the minor isophotal diameter (col.~7); position angle (PA) of the disk photometric 
major axis (col.~8); PA of the kinematic major axis of the stellar
velocity field, derived from this work (see \S~\ref{subsec:kin}, col.~9);
PA and semi-major axis length of the main stellar bar
(cols.~10 \& 11). References for the bar parameters are as follows: 1)
\cite{jun97}, 2) \cite{kna02}, 3) \cite{jog02b}. For IC\,1438, the inclination
estimate in Col. 8 is from \cite{gar02}, while no values are available
for the bar parameters.  \label{tab:sample} }
\end{table*}

\begin{deluxetable}{llllc}
\tablecaption{SINFONI observations\label{tab:obs}}
\tablewidth{0pt}
\tablehead{
\colhead{Galaxy} & 
\colhead{$t_{int}$}  &  
\colhead{OB}  &  
\colhead{Obs. date} &
\colhead{Seeing\tablenotemark{a} [$\as$] }   
} 
\startdata
	  &     & 216155  &    17.11.2005 & 0.6  \\
	  &     & 216157  &    13.11.2005 & 0.5  \\
NGC\,613  & 125 & 216158  &    23.10.2005 & 0.5  \\
	  &     & 216159  &    13.11.2005 & 0.5  \\
          &     & 216180  &    27.11.2005 &  0.6 \\
\hline
NGC\,1079 &  50 & 216161  &    17.11.2005 &  0.6 \\
	  &     & 216162  &    23.10.2005 & 0.5  \\
\hline
NGC\,1300 &  25 & 216168  &    24.1.2006 &  0.6 \\
\hline
	  &     & 216184  &    10.5.2006 & 0.7  \\
	  &     & 216187  &    15.5.2006 & 0.6  \\
NGC\,5248 & 125 & 216188  &    31.5.2006 & 0.7  \\
	  &     & 216189  &    1.6.2006 & 0.5 \\
	  &     & 216190  &    5.7.2006 & 0.5  \\
\hline
	  &     & 216147  &    22.10.2005 & 0.7  \\
	  &     & 216169  &    20.10.2005 & 0.6  \\
	  &     & 216170  &    19.10.2005 & 0.5  \\
IC\,1438  & 150 & 216171  &    19.10.2005 & 0.5  \\
	  &     & 216176  &    20.10.2005 & 0.6  \\
	  &     & 216278  &    22.10.2005 & 0.7  \\
\hline
\enddata 
\tablenotetext{a}{The atmospheric seeing was estimated from the FWHM of the
K-band point spread function, measured from standard star observations
during during the respective nights. These were normally scheduled 
within one hour from the galaxy exposures.}  
\end{deluxetable}

\begin{deluxetable}{lccc}
\tablecaption{
Summary of ring morphologies\label{tab:rings}}
\tablewidth{0pt}
\tablehead{
\colhead{Galaxy} & 
\colhead{$D$ [pc]}  &  
\colhead{PA[deg]} &
\colhead{$\epsilon$}    
} 
\startdata
NGC\,613  & 603   & 113  & 0.40     \\
NGC\,1079 & 238   &  72  & 0.29     \\
NGC\,1300 & 666   & 117  & 0.13     \\
NGC\,5248 & 338   & 102  & 0.39     \\
IC\,1438  & 1074  &  81  & 0.15     \\
\enddata 
\end{deluxetable}

\begin{deluxetable}{lcccc}
\tablecaption{Summary of emission line fluxes.\label{tab:flux}}
\tablewidth{0pt}
\tablehead{
\colhead{Galaxy} & 
\colhead{He I}  &  
\colhead{\brg} &
\colhead{\htwo}  &  
\colhead{\fetwo} 
} 
\startdata
NGC\,613     &  &  &  &   \\
\, - ring    &    46.5  &    136.5 & 80.5 & 184.2  \\
\, - nucleus &  $<$0.5  &   $<$1.1 & 6.1 &  11.8  \\
\hline 
NGC\,1079    &  &  &  &  \\
\, - ring    &  6.2    &  14.5    &  0.8    &  3.9 \\
\, - nucleus &  $<$0.2 &   1.5$^a$ &  $<$0.2  &  $<$0.2 \\
\hline 
NGC\,1300    &  &  &  &  \\
\, - ring    & 2.2 &   19.6 &  22.5 &  15.5 \\
\, - nucleus & $<$0.2 & $<$0.2 &  $<$1.1 &  $<$1.1 \\
\hline 
NGC\,5248    &  &  &  &  \\
\, - ring    &  1.2 &    4.5 &  8.4 & $<$0.2 \\
\, - nucleus &  $<$0.2 & $<$0.2 &  $<$1.0 & $<$0.2 \\
\hline 
IC\,1438     &  &  &  &  \\
\, - ring    &   10.2 & 13.8 &    5.2 &   10.6 \\
\, - nucleus & $<$0.2 &  $<$1.0 & $<$0.2 & $<$0.2 \\
\hline 
\enddata 
\tablecomments{All fluxes are in units
of $\rm 10^{-15}\,erg\,cm^{-2}\,s^{-1}$. Typical
uncertainties are about 20\% . }
\tablenotetext{a}{the central $1\as$ aperture is likely 
``contaminated'' by flux from the nuclear ring.}  
\end{deluxetable}

\clearpage

\clearpage

\clearpage

\clearpage

\clearpage

\clearpage

\clearpage

\clearpage

\clearpage

\clearpage

\clearpage

\clearpage
\end{document}